\documentclass[12pt]{article}
\usepackage{graphicx}
\usepackage{amsmath,epsfig}
\usepackage{longtable}
\usepackage{cite}
\usepackage{url}
\usepackage{multirow}
\usepackage{changes}

\newcommand{\aspi}{\frac{\hat\alpha_s}{\pi}}
\newcommand{\be}{\begin{equation}}
\newcommand{\ee}{\end{equation}}
\newcommand{\ba}{\begin{array}}
\newcommand{\ea}{\end{array}}
\newcommand{\bea}{\begin{eqnarray}}
\newcommand{\eea}{\end{eqnarray}}
\usepackage{color}

\begin{document}

\title{Charm Quark Mass with Calibrated Uncertainty}
\author{
Jens Erler 
  \\
  {\normalsize \em 
  Instituto de F\'isica, Universidad Nacional Aut\'onoma de M\'exico} 
  \\
  {\normalsize \em
  Apartado Postal 20--364, M\'exico D.F. 01000, M\'exico}
\\ \\
Pere Masjuan 
  \\
  {\normalsize \em
  Grup de F\'{\i}sica Te\`orica, Departament de F\'{\i}sica, 
  Universitat Aut\`onoma de Barcelona,}
  \\
  {\normalsize \em
  and Institut de F\'{\i}sica d'Altes Energies (IFAE), }
  \\
  {\normalsize \em
  The Barcelona Institute of Science and Technology (BIST), }
  \\
  {\normalsize \em
  Campus UAB, E-08193 Bellaterra (Barcelona), Spain}
\\ \\
and 
\\ \\
Hubert Spiesberger 
  \\
  {\normalsize \em 
  PRISMA Cluster of Excellence, Institut f\"ur Physik,}
  \\
  {\normalsize \em
  Johannes Gutenberg-Universit\"at, 55099 Mainz, Germany,}
  \\
  {\normalsize \em
  and Centre for Theoretical and Mathematical Physics and 
  Department of Physics,}
  \\
  {\normalsize \em
  University of Cape Town, Rondebosch 7700, South Africa}
}

\maketitle

\begin{abstract}
We determine the charm quark mass $\hat{m}_c$ from QCD sum rules of 
moments of the vector current correlator calculated in perturbative 
QCD at ${\cal O} (\hat\alpha_s^3)$. Only experimental data for the 
charm resonances below the continuum threshold are needed in our 
approach, while the continuum contribution is determined by requiring 
self-consistency between various sum rules. Existing data from the 
continuum region can then be used to bound the theoretic uncertainty.
Our result is $\hat{m}_c(\hat{m}_c) = 1272 \pm 8$~MeV for 
$\hat\alpha_s(M_Z) = 0.1182$.
\end{abstract}

\vfill
\quad MITP/16-087



\section{Introduction}

As a fundamental parameter of the Standard Model (SM) of particle 
physics that enters the calculation of many physical observables,
it is important to determine the charm quark mass as accurately 
as possible and by using several different methods. Moreover, it 
is one of the central goals of future high-energy lepton colliders 
to measure the charm quark Yukawa coupling to sub-percent precision 
in order to test the mass-coupling relation, as predicted in the 
minimal SM with a single Higgs doublet. Thus, the precision in the 
charm quark mass should at least match and ideally surpass that in 
the charm Yukawa coupling.

Most recent determinations~\cite{Erler:2014eya} of the $\overline 
{\rm MS}$ charm quark mass, $\hat{m}_c(\mu = \hat{m}_c)$, including 
all of the most precise ones, rely on either lattice simulations 
or variants of QCD sum rules~\cite{Novikov:1977dq,Shifman:1978bx, 
Shifman:1978by}. Both approaches are based on rigorous field 
theoretical principles and can be systematically improved.
Nevertheless, the resulting uncertainties are dominated by theory 
errors which are notoriously difficult to estimate and often subject 
of vigorous debate. 

The goal of this article is then to revisit the relativistic sum 
rule method with emphasis on the evaluation of the uncertainty and 
specifically to show that the overall error may also be constrained 
within our approach. To this end, we will adopt a strategy where 
the only exploited experimental information are the masses and 
electronic decay widths of the narrow resonances in the sub-continuum 
charm region, $J/\Psi$ and $\Psi(2S)$. Consistency between 
different QCD sum rules will be seen to suffice to constrain the 
continuum of charm pair production with good precision. For this 
procedure to work it is crucial to include alongside the first and 
higher moment sum rules also the zeroth moment, as the latter 
exhibits enhanced sensitivity to the continuum. Comparison with 
existing data on the $R$-ratio for hadronic relative to leptonic 
final states in $e^+ e^-$ annihilation will then serve as a control, 
providing an independent error estimate which we interpret as the 
error on the method and add it as an additional error contribution. 
In this way, we can show that the overall precision in $\hat{m}_c$ 
from relativistic sum rules is at the sub-percent level. 

Specifically, there are two sources of uncertainty which are 
particularly difficult to quantify, and therefore occasionally 
not even included in the error budget. One are condensate 
contributions within the operator product expansion (OPE) beyond 
the leading gluon condensate term. The other issue is quark-hadron 
duality: Formally, the sum rule is based on the optical theorem, 
where the real part of the correlator is evaluated theoretically 
with quark and gluon degrees of freedom, while the imaginary part 
involves as {\em dual} degrees of freedom the hadrons observed in 
experiments. While such a global change of variables is highly 
non-trivial, it is innocuous as long as one does not mix these 
descriptions on either side of the sum rule. As far as the 
evaluation of the imaginary part is concerned, one is forced, 
however, to switch at some specific point in the squared energy 
$s$ from experimental data to perturbative QCD since in practice 
data are necessarily constrained to a finite region while the 
upper integration limit is unbounded. In this way one has to rely 
on local quark-hadron duality which -- at least as a matter of 
principle -- is unjustified. Still, as long as $s$ is large 
enough, one introduces little additional uncertainty. Now, the 
largest value of $\sqrt{s}$ for which data of the total hadronic 
cross section in  $e^+ e^-$ annihilation are available (see the 
data points in Figure~\ref{Rcan} in Section~\ref{Sec:data}) is 5 GeV, 
and beyond $\sqrt{s} = 4.6$ GeV data points are scarce and have very 
large errors. Around this energy there is still considerable 
fluctuations in the measured cross-section, shedding some doubt 
on the applicability of local quark-hadron duality even in practice. 
One of the features of our work is that it merely relies on 
quark-hadron duality in a finite region, namely between the 
$\psi(2S)$ resonance and the continuum threshold. While this 
is still not rigorously justified, it should largely mitigate 
the aforementioned problem. Furthermore, by using continuum data 
only as a calibration of the uncertainty, we control the error 
associated with the necessary deviation from strict global 
quark-hadron duality.

The paper is organized as follows. In the next section we 
review the formalism following Ref.~\cite{Erler:2002bu} and 
describe how self-consistency between moments of the current 
correlator allows us to constrain the continuum region and 
determine a precise value of the charm quark mass. This section 
will already contain our main result, $\hat{m}_c(\hat{m}_c) = 
1272 \pm 8$~MeV for $\hat\alpha_s(M_Z) = 0.1182$. In 
Section~\ref{Sec:data} we add a detailed discussion of the 
influence of the continuum region using experimental data and 
confirm the validity of the result of Section \ref{Sec:formalism}. 
Section~\ref{Sec:errors} offers details of a more general fit 
procedure where parameter uncertainties can be taken into account 
in a more systematic way, and we also include a comparison with 
earlier results. We summarize in Section~\ref{Sec:conclusions}.


\section{Formalism and charm-mass determination}
\label{Sec:formalism}

We consider the transverse part of the correlator $\hat\Pi_q(t)$ 
of two heavy-quark vector currents where the caret indicates 
$\overline{\rm MS}$ subtraction. $\hat\Pi_q(t)$ can be calculated 
in perturbative QCD (pQCD) order by order and obeys the subtracted 
dispersion relation~\cite{Chetyrkin:1994js}
\be
  12 \pi^2 \frac{\hat\Pi_q (0) - \hat\Pi_q (-t)}{t} =
  \int_{4 \hat m_q^2}^\infty \frac{{\rm d} s}{s} 
  \frac{R_q(s)}{s + t} \ ,
\label{SR}
\ee
where $R_q(s) = 12 \pi \mbox{Im} \hat\Pi_q(s)$, and $\hat m_q = 
\hat m_q (\hat m_q)$ is the mass of the heavy quark. In the limit 
$t\rightarrow 0$, Eq.~(\ref{SR}) coincides with the first moment, 
${\cal M}_1$, of $\hat\Pi_q(t)$. In general, there is a sum rule 
for each higher moment as well~\cite{Novikov:1977dq,Shifman:1978bx, 
Shifman:1978by},
\be
  {\cal M}_n 
  := 
  \left.\frac{12\pi^2}{n !} \frac{d^n}{d t^n} 
  \hat\Pi_q(t) \right|_{t=0} 
  = 
  \int_{4 \hat m_q^2}^\infty \frac{{\rm d} s}{s^{n+1}} R_q(s).
\label{SRder}
\ee
Taking the opposite limit of Eq.~(\ref{SR}), i.e.\ $t \rightarrow 
\infty$, multiplying with $t$ and properly regularizing, one can 
also define a sum rule for the $0^{th}$ moment~\cite{Erler:2002bu}. 
${\cal M}_0$ is then obtained from the dispersion relation for the 
difference $\hat\Pi_q (0) - \hat\Pi_q (-t)$ where the (unphysical) 
constant $\hat\Pi_q (0)$ is subtracted, tantamount to the definition 
of higher moments as derivatives of $\hat\Pi_q (-t)$. At a given 
order in pQCD, the required regularization can be obtained by 
subtracting the zero-mass limit of $R_{q}(s)$, which we write as 
$3 Q_q^2 \lambda_1^q(s)$ with $Q_q$ the quark charge. $\lambda_1^q(s)$ 
is known up to ${\cal O}(\hat\alpha_s^4)$, but we will only need the 
third-order expression~\cite{Chetyrkin:2000zk},
\bea \label{lambda1}
  \lambda^q_1(s) 
  &=& 1 
  + \frac{\hat \alpha_s}{\pi}
  + \frac{\hat \alpha_s^2}{\pi^2} \left[ \frac{365}{24} - 11 \zeta(3)
     + n_q \left( \frac{2}{3} \zeta(3) - \frac{11}{12} \right)  \right] 
  \\ \nonumber
  &+& \frac{\hat \alpha_s^3}{\pi^3} 
     \left[\frac{87029}{288} - \frac{121}{8} \zeta(2) 
     - \frac{1103}{4} \zeta(3) + \frac{275}{6}\zeta(5) 
     \right. 
  \\ \nonumber 
  &+&  \left. n_q \left(- \frac{7847}{216} + \frac{11}{6} \zeta(2) 
  + \frac{262}{9} \zeta(3) - \frac{25}{9}\zeta(5) \right) 
  + n_q^2 \left(\frac{151}{162} - \frac{\zeta(2)}{18} 
  - \frac{19}{27} \zeta(3) \right)  \right],
\eea
where $\hat \alpha_s = \hat \alpha_s (\sqrt s)$, $n_q = n_l + 1$ 
and $n_l$ is the number of light flavors (taken as massless), 
{\em i.e.}, quarks with masses below the heavy quark under 
consideration. 

\begin{table}[t]
\begin{center}
\begin{tabular}{|c|r|r|}
\hline
\rule[-3mm]{0mm}{9mm}
$R$ & $M_R$ [GeV] & $\Gamma_R^e$ [keV] \\
\hline
 \rule[-2mm]{0mm}{7mm}
 $J/\Psi$ & 3.096916 & 5.55(14)  \\
 \rule[-2mm]{0mm}{6mm}
 $\Psi(2S)$ & 3.686109 & 2.36\phantom1(4)  \\
\hline
\end{tabular}
\end{center}
\caption{
Resonance data~\cite{Agashe:2014kda} used in the analysis. 
The uncertainties from the resonance masses are negligible 
for our purpose.}
\label{ResPDG}
\end{table}

By the optical theorem, $R_{q}(s)$ can be related to the measurable 
cross section for heavy-quark production in $e^+e^-$ annihilation. 
Below the threshold for continuum heavy-flavor production, 
the cross section is determined by a small number of narrow 
Breit-Wigner resonances \cite{Novikov:1977dq}, approximated by 
$\delta$-functions, 
\be
  R_q^{\rm Res}(s) = 
  \frac{9\pi}{\alpha_{\rm em}^2(M_R)} M_R \Gamma_R^e\delta(s-M_R^2) .
\label{Rres}
\ee
The masses $M_R$ and electronic widths $\Gamma^e_R$ of the resonances 
\cite{Agashe:2014kda} needed for the determination of the charm mass 
are listed in Table~\ref{ResPDG} and $\alpha_{\rm em}(M_R)$ is the 
running fine structure constant at the resonance\footnote{The values 
for $\alpha_{\rm em}(M_R)$ were determined with help of the 
program {\tt hadr5n12} \cite{jegerlehner-url}.}.

We invoke global quark-hadron duality, and we assume that continuum 
production can be described on average by the simple ansatz 
\cite{Erler:2002bu},
\be 
  R_q^{\rm cont}(s)
  = 3 Q^2_q \lambda^q_1 (s) 
  \sqrt{1 - \frac{4\, \hat{m}_q^2 (2 M)}{s^\prime}} 
  \left[ 1 + \lambda^q_3 \frac{2\, \hat{m}_q^2(2 M)}{s^\prime} 
  \right],
\label{ansatz}
\ee
where $s' := s + 4(\hat{m}_q^2(2M) - M^2)$, and $M$ is taken as the 
mass of the lightest pseudoscalar charmed meson, {\em i.e.}, $M = M_{D^0} = 1864.84$ MeV~\cite{Agashe:2014kda}. 
$\lambda^q_3$ is a free parameter to be determined. $R_q^{\rm cont}(s)$ 
interpolates smoothly between the threshold and the onset of open 
heavy-quark pair production and coincides asymptotically with the 
prediction of pQCD for massless quarks. 

\begin{table}[t]
\begin{center}
\begin{tabular}{|c|c|c|c|c|}
\hline
\rule[-3mm]{0mm}{9mm}
$n$
& $C_n^{(0)}$
& $C_n^{(1)}$
& $C_n^{(2)}$
& $C_n^{(3)}$
\\[5pt]
\hline
\rule[-3mm]{-4pt}{9mm}
$1$ 
& $1.0667$ 
& $2.5547$ 
& $2.4967$
& $-5.6404$ 
\\[5pt]
$2$ 
& $0.4571$ 
& $1.1096$ 
& $2.7770$
& $-3.4937$ 
\\[5pt]
$3$ 
&$0.2709$ 
& $0.5194$ 
& $1.6388$
& $-2.8395$ 
\\[5pt]
$4$ 
& $0.1847$ 
& $0.2031$ 
& $0.7956$
& $-3.722 \pm 0.500$
\\[5pt]
$5$ 
& $0.1364$ 
& $0.0106$ 
& $0.2781$
& $-4.425 \pm 1.200$
\\
\hline
\end{tabular}
\end{center}
\caption{
The coefficients $C_n^{(i)}$ for the perturbative expansion of 
the QCD moments, Eq.\ (\ref{Mth}, \ref{Cnth}), as used in this 
work. The first line for $n=$1 is taken from Ref.\ 
\cite{Chetyrkin:2006xg}, the coefficients for $n=2$, 3 from 
\cite{Maier:2009fz} and the last ones for $n=4$, 5 from Ref.\ 
\cite{Greynat:2011zp}.
}
\label{Tab:Cni}
\end{table}

Using the results of Refs.~\cite{Chetyrkin:1996cf,Chetyrkin:1997un} 
and $R_q(s) = \sum\limits_{\rm resonances} R_q^{\rm Res}(s) + 
R_q^{\rm cont}(s)$, the sum rule for ${\cal M}_0$ obtained from 
Eq.~(\ref{SR}) in the limit $t \rightarrow \infty$ and dividing 
by $3Q_q^2$ reads explicitly,
\bea \nonumber
  && 
  \sum\limits_{\rm resonances} 
  \frac{9\pi\Gamma^e_R}{3 Q_q^2 M_R \alpha_{em}^2 (M_R)} 
+ \int\limits_{4 M^2}^\infty \frac{{\rm d} s}{s}  \left(
   \frac{R_q^{\rm cont}(s)}{3 Q_q^2} -\lambda_1^q(s)\right)
  - \int\limits_{\hat{m}_q^2}^{4M^2}
   \frac{{\rm d} s}{s} \lambda^q_1 (s) 
  \nonumber \\ &&
  = - \frac{5}{3} + \aspi \left[ 4 \zeta(3) - \frac{7}{2} \right] 
\label{SR0}
  \\  && \quad
  + \left(\aspi \right)^2  
   \left[ \frac{2429}{48} \zeta(3) - \frac{25}{3} \zeta(5) 
   - \frac{2543}{48} 
   + n_q \left( \frac{677}{216} - \frac{19}{9} \zeta(3) \right) 
   \right] + \left(\aspi \right)^3 A_3\ ,
  \nonumber \\ 
  &&
  = -1.667 + 1.308\, \aspi + 1.595 \left(\aspi\right)^2 
  - 8.427 \left(\aspi\right)^3 \, ,
  \nonumber
\eea
where $\hat\alpha_s = \hat \alpha_s(\hat m_q)$ and the third-order 
coefficient $A_3$ is available in numerical form~\cite{Kiyo:2009gb, 
Hoang:2008qy},
\be
A_3 = -9.863 + 0.399 \, n_q - 0.010 \, n_q^2\ .
\ee
In the last line of Eq.\ (\ref{SR0}) we show numerical values 
for the case of charm quarks, i.e.\ $n_q=4$, exhibiting a marked 
breakdown of convergence of the perturbative expansion. 
Notice that the continuum $R_q^{\rm cont}(s)$ contributes with the 
lower integration limit $4M^2$, while the subtraction term 
$\lambda_1^q(s)$ is integrated starting from $\hat{m}_q^2$.  
We reiterate that this definition of ${\cal M}_0$ does not 
involve the unphysical constant $\hat\Pi_q (0)$ separately, but 
only the difference $\hat\Pi_q (0) - \hat\Pi_q (-t)$. 

\begin{table}[t]
\begin{center}
\begin{tabular}{|c|c|c|}
\hline
\rule[-3mm]{0mm}{9mm}
$n$
& $a_n$
& $b_n$
\\[5pt]
\hline
\rule[-3mm]{-4pt}{9mm}
$0$ 
& $-\frac{2}{15}$ 
& $\frac{605}{162}$ 
\\[5pt]
$1$ 
& $-\frac{32}{105}$ 
& $\frac{32 099}{12 960}$ 
\\[5pt]
$2$ 
& $-\frac{32}{63}$ 
& $\frac{59}{56}$ 
\\[5pt]
$3$ 
& $-\frac{512}{693}$ 
& $-\frac{20 579}{42 525}$ 
\\[5pt]
$4$ 
& $-\frac{1280}{1287}$ 
& $-\frac{100 360 567}{47 628 000}$ 
\\[5pt]
$5$ 
& $-\frac{8192}{6435}$ 
& $-\frac{459 884 251}{121 080 960}$ \\[5pt]
\hline
\end{tabular}
\end{center}
\caption{
The coefficients $a_n$ and $b_n$ needed to calculate the 
contribution from the dimension-4 gluon condensate to the 
QCD moments, Eq.\ (\ref{Mncond}). The numbers are taken from 
Refs.~\cite{Broadhurst:1994qj,Kuhn:2007vp,Chetyrkin:2010ic}.
}
\label{Tab:anbn}
\end{table}

Eq.~(\ref{SR0}) contains two unknowns, the quark mass 
$\hat m_q(\hat m_q)$ and the parameter $\lambda_3^q$ entering in 
our prescription for $R_q^{\rm cont}(s)$. The zeroth sum rule is 
the most sensitive to the continuum region and therefore to 
$\lambda_3^q$, while the other moments mostly determine the 
quark mass. 

Theory predictions for the higher moments in perturbative QCD can 
be cast into the form 
\be
  {\cal M}_n^{\rm pQCD}  = 
  \frac{9}{4} Q_q^2  
  \left( \frac{1}{2\hat{m}_q(\hat{m}_q)} \right)^{2n}  \hat{C}_n 
\label{Mth}
\ee
with
\be
  \hat{C}_n = 
  C_n^{(0)} 
  + \left(\aspi + \frac{3 Q_q^2 \alpha_{\rm em}}{4 \pi}\right)C_n^{(1)} 
  + \left(\aspi \right)^2C_n^{(2)} 
  + \left(\aspi \right)^3C_n^{(3)} 
  + {\cal O}(\hat\alpha_s^4) \, .
\label{Cnth}
\ee
The $C_n^{(i)}$ are known up to ${\cal O}(\hat\alpha_s^3)$ for 
$n \leq 3$ \cite{Chetyrkin:2006xg,Boughezal:2006px,Maier:2008he, 
Maier:2009fz}, and up to ${\cal O}(\hat\alpha_s^2)$ for the 
rest~\cite{Chetyrkin:1997mb,Maier:2007yn}. For the convenience 
of the reader we collect the numerical values of the coefficients 
$C_n^{(i)}$ required for our analysis in Table~\ref{Tab:Cni}. Since 
we evaluate the moments up to ${\cal O} (\hat\alpha_s^3)$ we use 
the predictions for $n>3$ provided in Ref.~\cite{Greynat:2011zp}. 
The QED contribution to the QCD moments proportional to 
$\alpha_{\rm em} / \pi$ is very small\footnote{Not taking into 
account the QED contribution would decrease the value of 
$\hat m_c(\hat m_c)$ by about $0.5$ MeV.} and it is sufficient 
to include the first-order term. 

\begin{table}[t]
\begin{center}
\begin{tabular}{|c|c|r|r|r|}
\hline
\rule[-3mm]{0mm}{9mm}
$n$ & Resonances & Continuum & Total & Theory \hspace{8pt} \\
\hline
 \rule[-2mm]{0mm}{7mm}
0 & 1.231 (24) &$-3.229 (+28)(43)(1)$ &$-1.999(56)$ & {\bf Input} (11) \\
 \rule[-2mm]{0mm}{6mm}
1 & 1.184 (24) & $0.966 (+11)(17)(4)$ & $2.150 (33)$ & 2.169(16) \\
 \rule[-2mm]{0mm}{6mm}
2 & 1.161 (25) & $0.336 (+5)(8)(9)$ & $1.497 (28)$ & {\bf Input} (25) \\
 \rule[-2mm]{0mm}{6mm}
3 & 1.157 (26) & $0.165(+3)(4)(16)$ & $1.322 (31)$ & 1.301(39) \\
 \rule[-2mm]{0mm}{6mm}
4 & 1.167 (27) & $0.103 (+2)(2)(26)$ & $1.270 (38)$ & 1.220(60) \\
 \rule[-2mm]{0mm}{6mm}
5 & 1.188 (28) & $0.080 (+1)(1)(38)$ & $1.268 (47)$ & 1.175(95) \\
\hline
\end{tabular} 
\end{center}
\caption{
Results for the lowest moments, ${\cal M}_n$, defined in Eq.~(\ref{SR0}) 
for $n = 0$ (multiplied by $3Q_q^2$) and Eq.~(\ref{SRder}) for $n \geq 
1$ and including the contribution from the gluon condensate, Eq.\ 
(\ref{Mncond}) (multiplied by $10^n\, \mbox{GeV}^{2n}$). The first 
error for the continuum part is obtained from shifting the central 
value $\lambda_3^c = 1.23$ (determined from the 0th and the 2nd moment) 
to $\lambda_3^{c,{\rm exp}} = 1.34(17)$ (determined from fixing the 0th 
moment by experimental data, see next section). The second error is 
propagated from the experimental uncertainties of data in the 
threshold region, $\Delta \lambda_3^{c,{\rm exp}} = \pm 0.17$. The 
third error is due to the uncertainty of the gluon condensate. The 
errors for the sum of the resonance and continuum parts (denoted 
'Total' in column 4) combines all errors in quadrature. The last 
column shows the theoretical predictions using Eq.~(\ref{Mth}) 
for $\hat{m}_c (\hat{m}_c) = 1.272$ GeV and $\hat\alpha_s(M_Z) = 
0.1182$ \cite{alphas} with the truncation error determined 
from Eq.~(\ref{truncerror}). 
}
\label{moments}
\end{table}

In general, vacuum expectation values of higher-dimensional 
operators in the OPE contribute to the moments of the current 
correlator as well. These condensates may be important for a 
high-precision determination of heavy-quark masses, in particular 
in the case of the charm quark. The leading term involves the 
dimension-4 gluon condensate \cite{Novikov:1977dq}, 
\begin{equation}
  {\cal M}^{\rm cond}_n  
  = 
  \frac{12 \pi^2 Q_q^2}{(4 \hat m_q^2)^{n+2}} 
  \left\langle \frac{\hat\alpha_s}{\pi} G^2 \right\rangle \, 
  a_n \left(1 + \frac{\hat\alpha_s(\hat m_q)}{\pi} b_n\right) \, .
\label{Mncond}
\end{equation}
The coefficients $a_n$ and $b_n$ can be found in 
Refs.~\cite{Broadhurst:1994qj,Kuhn:2007vp,Chetyrkin:2010ic} and 
are collected in Table~\ref{Tab:anbn}. We will use $C_G^{\rm exp} 
:= \langle \frac{\hat\alpha_s}{\pi} G^2 \rangle^{\rm exp} = 
0.005$~GeV$^4$ as our central value and assume a 100\% uncertainty, 
$\Delta_G = 0.005$~GeV$^4$, taken from the recent 
analysis~\cite{Dominguez:2014fua}. 

\begin{table}[t]
\begin{center}
\begin{tabular}{|c|c|c|c|}
\hline
\rule[-3mm]{0mm}{9mm}
$n$ & 
$\frac{\Delta {\cal M}_n^{(2)}}{ \left|{\cal M}_n^{(2)}\right|}$ 
& $\frac{\Delta {\cal M}_n^{(3)}}{ \left|{\cal M}_n^{(3)}\right|}$ 
& $\frac{\Delta {\cal M}_n^{(4)}}{ \left|{\cal M}_n^{(3)}\right|}$ 
\\[4mm]
\hline
0 & 1.88 & 3.03 &  1.11
\\
1 & 2.14 & 2.84 &  1.04
\\
2 & 1.92 & 4.58 &  1.67
\\
3 & 3.25 & 5.63 &  2.06
\\
4 & 6.70 & 4.30 &  1.57
\\
5 & 19.18& 3.62 &  1.32
\\
\hline
\end{tabular}
\end{center}
\caption{
Ratios of the truncation errors $\Delta {\cal M}_n^{(k)}$ from 
Eq.~(\ref{truncerror}) and the known contribution to the moments 
${\cal M}_n^{(k)}$ at order $O(\hat\alpha_s^k)$ (see Eq.\ 
(\ref{Mth})). In the last column, $\hat{m}_c (\hat{m}_c) = 
1.272$~GeV has been used. The ratios are often much greater 
than unity, showing that our estimated truncation errors are 
conservative. 
}
\label{Tab:ErrorThMom}
\end{table}

In the following we will use this formalism for a determination 
of the charm quark mass and specify the notation accordingly, 
using the index $c$ instead of $q$ where appropriate. One can use  
\be
  {\cal M}_n^{\rm pQCD} +   {\cal M}^{\rm cond}_n 
  = 
  \int_{4 \hat m_c^2}^\infty \frac{{\rm d} s}{s^{n+1}} R_c(s)   
  = 
  \int_{4 \hat m_c^2}^\infty \frac{{\rm d} s}{s^{n+1}} 
  [R_c^{\rm Res}(s) + R_c^{\rm cont}(s)]
 \label{Mnth}
\ee
for two different $n$ to determine values for the heavy quark mass 
$\hat{m}_c(\hat{m}_c)$ and the constant $\lambda^c_3$. The other 
moments are then fixed and can be used to check the consistency of 
our approach. No experimental data other than the resonance parameters 
in Table~\ref{ResPDG} are necessary. Numerical results choosing 
$n=0$ and $n=2$ in Eq.~(\ref{Mnth}) are shown in Table~\ref{moments}. 
We show separately the contributions from the narrow resonances 
(column 2) and from the continuum part (column 3), evaluated using 
$R_c^{\rm cont}(s)$ in Eq.\ (\ref{ansatz}), including the condensate 
contribution from Eq.~(\ref{Mncond}). The sum of these two 
contributions (column 4) can be compared with the theory prediction 
(column~5) from Eq.\ (\ref{Mth}). The second column in 
Table~\ref{moments} accounts for the narrow vector resonances below 
the heavy-quark threshold, in the charm sector the first two 
charmonium resonances, $J/\Psi$ and $\Psi(2S)$. The higher 
resonances are included in the continuum. The errors for the 
resonance contributions shown in Table~\ref{moments} are exclusively 
determined by the uncertainty of the electronic widths from 
Table~\ref{ResPDG}, taken to be 
uncorrelated\footnote{The error coming from the electronic partial 
   widths is completely dominated by the $J/\Psi$; the $\Psi(2S)$ 
   contributes only about $\pm 0.002$ to the error of the moments. 
   All errors are combined in quadrature. Had we assumed them to 
   be completely correlated, the errors would have been slightly 
   increased, by about 0.004, 0.004, 0.003, 0.002, 0.002 for the 
   first to the fifth moment, respectively.
}. 

\begin{figure}[t]
\begin{center}
\includegraphics[width=0.8\textwidth]{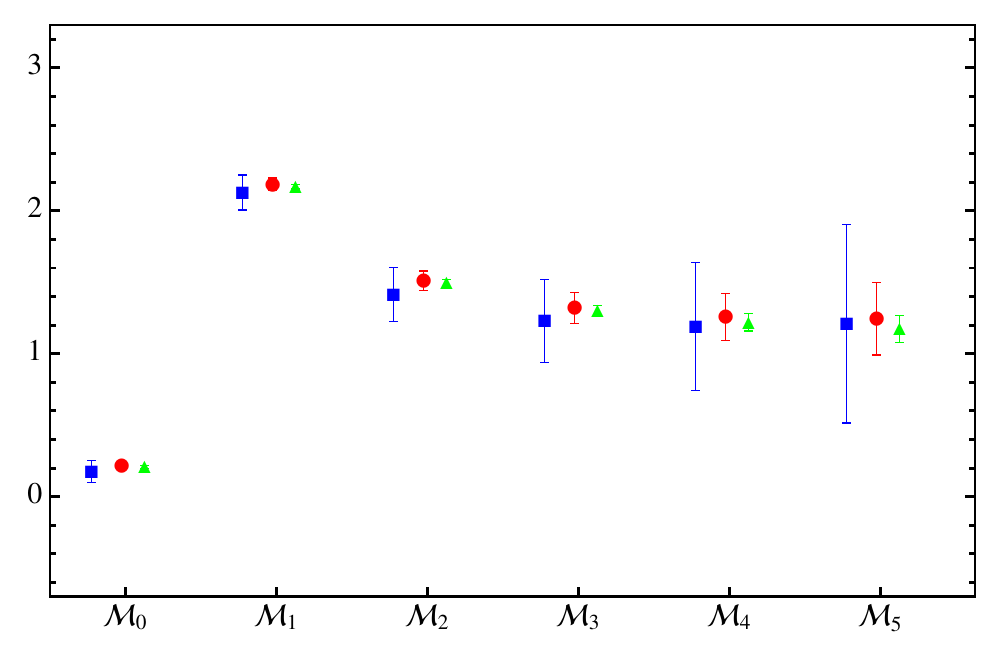}
\caption{
Theoretical moments, Eq.~(\ref{Mth}), (multiplied by $10^n\, 
\mbox{GeV}^{2n}$) for the charm quark using $\hat{m}_c(\hat{m}_c) = 
1.272$ GeV at different orders of $\hat\alpha_s$. Blue squares show 
results of Eq.~(\ref{Mth}) up to ${\cal O}(\hat\alpha_s)$, red 
circles up to ${\cal O}(\hat\alpha_s^2)$, and green triangles up 
to ${\cal O}(\hat\alpha_s^3)$. The error bars show the truncation 
errors given in Eq.~(\ref{truncerror}) at the given order. 
}
\label{cmmoments}
\end{center}
\end{figure}

\begin{table}[t!]
\begin{center}
\begin{tabular}{|c|ccccc|}
\hline
\rule[-3mm]{0mm}{9mm}
$\Delta \hat{m}_c(\hat{m}_c)$
& $({\cal M}_0, {\cal M}_1)$ 
& {\boldmath$({\cal M}_0, {\cal M}_2)$} 
& $({\cal M}_0, {\cal M}_3)$ 
& $({\cal M}_0, {\cal M}_4)$ 
& $({\cal M}_0, {\cal M}_5)$ 
\\
\hline
\rule[-2mm]{0mm}{7mm}
$\hat{m}_c(\hat{m}_c)$ 
& 1280.9 
& {\bf 1272.4} 
& 1269.1 
& 1265.8 
& 1262.2 
\\ 
\rule[-2mm]{0mm}{7mm}
$\lambda_3^c$ 
& 1.154 
& {\bf 1.230} 
& 1.262 
& 1.291 
& 1.323 
\\
\rule[-2mm]{0mm}{7mm}
$\lambda_3^{c,{\rm exp}}$ 
& 1.35(17) 
& {\bf 1.34(17)}
& 1.34(17) 
& 1.33(17)
& 1.32(17)
\\
\hline
\rule[-2mm]{0mm}{7mm}
Resonances 
& 5.8 
& {\bf 4.5} 
& 3.9 
& 3.3 
& 2.8 
\\
\rule[-2mm]{0mm}{7mm}
Truncation
& 6.3
& {\bf 5.9}
& 7.2
& 8.9
& 10.5
\\ 
\rule[-2mm]{0mm}{7mm}
$\lambda_3^c - \lambda_3^{c,{\rm exp}}$ 
& +6.4 
& {\bf +1.5} 
& +0.3 
& +0.1 
& +0.1 
\\
\rule[-2mm]{0mm}{7mm}
$\Delta \lambda_3^{c,{\rm exp}}$ 
& 4.7 
& {\bf 1.7} 
& 0.7 
& 0.3 
& 0.2 
\\
\rule[-2mm]{0mm}{7mm}
$10^3\times \Delta_G$
& $-0.25 \Delta_G$ 
& {\boldmath $-0.37 \Delta_G$} 
& $-0.54 \Delta_G$ 
& $-0.73 \Delta_G$ 
& $-0.88 \Delta_G$ 
\\
\rule[-2mm]{0mm}{7mm}
& $(-1.3)$
& ${\bf (-1.9)}$
& $(-2.7)$
& $(-3.7)$
& $(-4.4)$
\\
\hline
\rule[-3mm]{0mm}{9mm}
Total 
& $\pm 11.7$ 
& {\boldmath $\pm 8.0$} 
& $\pm 8.7$ 
& $\pm 10.2$ 
& $\pm 11.7$ 
\\
\hline
\rule[-2mm]{0mm}{7mm}
$10^3\times\Delta \hat\alpha_s(M_z)$ 
& $+3.6 \Delta \hat\alpha_s$ 
& {\boldmath $+2.6 \Delta \hat\alpha_s$} 
& $+1.6 \Delta \hat\alpha_s$ 
& $+0.6 \Delta \hat\alpha_s$ 
& $-0.4 \Delta \hat\alpha_s$ 
\\
\rule[-2mm]{0mm}{7mm}
Electroweak fit 
& (+5.8)
& {\bf (+4.2)}
& (+2.6)
& (+1.0)
& ($-0.6$)
\\
\rule[-2mm]{0mm}{7mm}
Lattice 
& (+4.3)
& {\bf (+3.1)}
& (+1.9)
& (+0.7)
& ($-0.5$)
\\
\hline
\end{tabular}
\end{center}
\caption{
Values and breakdown of the uncertainties of $\hat{m}_c(\hat{m}_c)$ 
(in MeV) and $\lambda_3^c$ determined from different pairs of 
moments. The line denoted 'Total' gives the quadratic sum of the 
errors from $\lambda_3^c$, the resonances, the gluon condensate 
and the pQCD truncation. Numerical values for the uncertainties 
from $\hat\alpha_s$ and $C_G = \langle \frac{\hat\alpha_s}{\pi} 
G^2 \rangle$ (in units of GeV$^4$) are shown in separate lines. 
In the line labeled 'Electroweak fit' we use $\Delta \hat\alpha_s(M_z) 
= 0.0016$ \cite{alphas}, in the last line denoted 'Lattice', we 
use $\Delta \hat\alpha_s(M_z) = 0.0012$ \cite{Agashe:2014kda}. 
See Figure~\ref{Fig:cmass} for a graphical representation.
}
\label{Tab:MomentsBudget}
\end{table}

\begin{figure}[t]
\begin{center}
\includegraphics[width=0.8\textwidth]{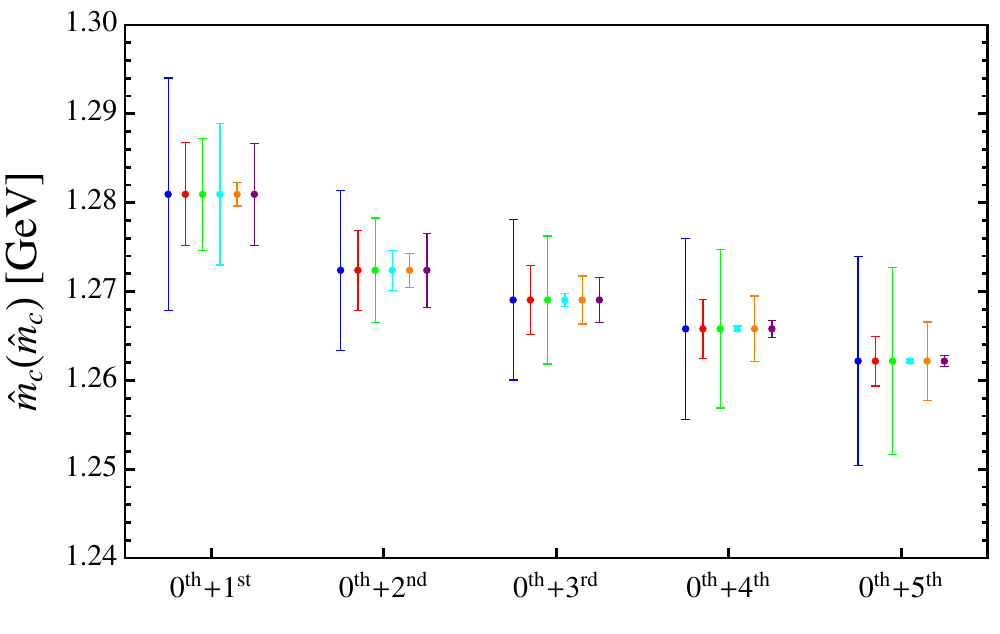}
\caption{
$\hat{m}_c(\hat{m}_c)$ using different combinations of moments and 
the corresponding uncertianties. Blue represents the full error, 
red is the one from the resonance region, green from the truncation 
errors of the theoretical moments, cyan from the error in
$\lambda_3^c$ (which is the symmetrized combination of the shift 
and the experimental error on $\lambda_3^c$), orange from the gluon 
condensate and purple from the uncertainty induced by 
$\Delta \hat\alpha_s(M_Z) = 0.0016$. Notice, that all determinations 
are mutually consistent within our
error estimates.}
\label{Fig:cmass}
\end{center}
\end{figure}

\begin{figure}[t!]
\begin{center}
\includegraphics[width=0.8\textwidth]{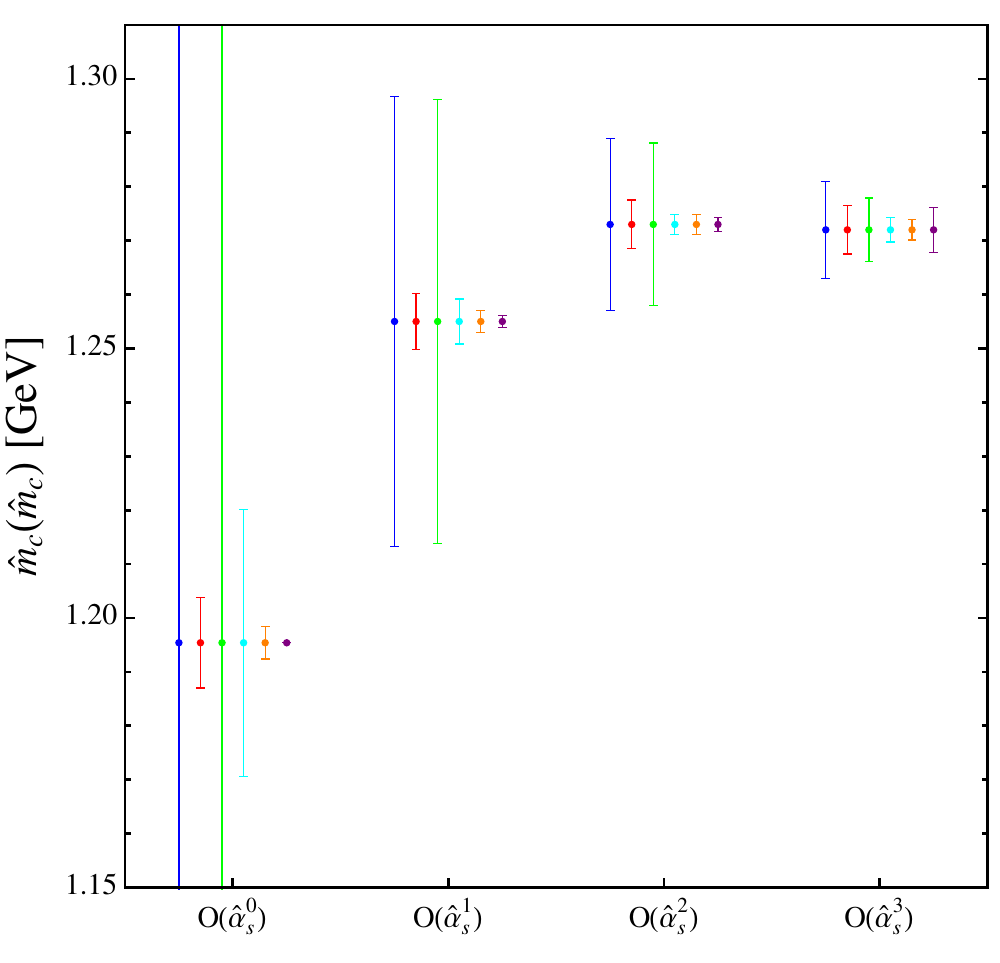}
\caption{
Error budget for $\hat{m}_c(\hat{m}_c)$ determined from the 
combination of ${\cal M}_0$ and ${\cal M}_2$ at different orders 
in $\hat\alpha_s$. The color coding is as in Figure~\ref{Fig:cmass}.}
\label{Fig:cmass-an}
\end{center}
\end{figure}

In order to determine an error for the continuum contributions 
(column 3) we proceed in the following way: we compare 
experimental data with the zeroth moment in the restricted energy 
range of the threshold region, $2 M_{D^0} \leq \sqrt{s} \leq 4.8$~GeV, 
to obtain an {\em experimental} value for $\lambda_3^c$, denoted 
$\lambda_3^{c,{\rm exp}}$ (details will be described below in 
Section~\ref{Sec:data}). Here we fix $\hat{m}_c(\hat{m}_c) = 1.272$~GeV. 
Then we can also determine an error, $\Delta \lambda_3^{c,{\rm exp}}$, 
from the experimental uncertainty of the data in this threshold 
region. The shift in the moments resulting from the different 
values for $\lambda_3^{c}$ (either from two moments combined with 
resonance data only, or from the comparison of the 0th moment with 
continuum data in the threshold region) turns out to be small. 
Strictly speaking this shift is a one-sided error, but to be 
conservative we include it as an additional error in the results 
of Table~\ref{moments}. Note that non-perturbative effects not 
included in our expressions for the moments, as for example 
omitted higher-dimensional condensates, would become visible 
in the comparison of the values of $\lambda_3^{c,{\rm exp}}$ 
determined from data with $\lambda_3^{c}$ obtained from moments. 
Since we include this difference as a contribution to the uncertainty 
budget of the charm mass, we do not include an additional uncertainty 
due to the truncation of the OPE. More details on our determination 
of $\lambda_3^{c,{\rm exp}}$ are given in the next section.

Finally, we assign a truncation error to the theory prediction of 
the moments. Following the method proposed in Ref.~\cite{Erler:2002bu} 
we consider the largest group theoretical factor in the next 
uncalculated perturbative order as a way to estimate these errors, 
\be
  \Delta {\cal M}_n^{(i)} 
  = \pm Q_q^2 
  N_C C_F C_A^{i-1} \, 
  \left(\frac{\hat\alpha_s (\hat{m}_q)}{\pi}\right)^i
  \left( \frac{1}{2 \hat{m}_q(\hat{m}_q)} \right)^{2n} 
  \, ,
\label{truncerror}
\ee
($N_C = C_A = 3$, $C_F = 4/3$). At order ${\cal O}(\hat\alpha_s^4)$, 
this corresponds to an uncertainty of $\pm 48 (\hat \alpha_s/\pi)^4$ 
for $\hat{C}_n^{(4)}$ in Eq.\ (\ref{Mth}). Applying this 
prescription to the known ${\cal O}(\hat\alpha_s^3)$ terms,
we observe that the resulting errors are quite conservative: 
the error estimate from this approach for $\hat{m}_c(\hat{m}_c)$ 
is more conservative by a factor given in the last column of 
Table~\ref{Tab:ErrorThMom} compared to the alternative to base 
the error estimate on the last known term in the perturbative 
expansion.  We have convinced ourselves that this carries over 
to the errors for the charm mass: the above prescription is more 
conservative than the often used alternative approach to vary 
the renormalization scale within a conventional factor of four. 

For the moments with $n>3$ (which, however,  will not enter
into our final result) taken from Ref.~\cite{Greynat:2011zp} 
we have to include additional uncertainties specific to the 
method used to obtain predictions for ${\cal M}_n$. These 
errors are small, but included for completeness. An overview 
of our theory errors for the moments up to $n=5$ is shown in 
Figure~\ref{cmmoments}.

The charm mass and the continuum parameter $\lambda_3^c$ can, 
in principle, be determined from any combination of two moments. 
The zeroth moment, however, provides the highest sensitivity. 
The results for combinations of the zeroth with one higher moment 
are summarized in Table~\ref{Tab:MomentsBudget}, including the 
breakdown of the uncertainties due to their different sources. We 
include the difference between the two possibilities to determine 
$\lambda_3^c$ as described above as an estimate of the error due 
to the method, in particular the truncation of the OPE. In 
Figure~\ref{Fig:cmass-an} we also show the error budget at different 
orders of $\hat \alpha_s$ as obtained from the combination of 
the moments ${\cal M}_0$ and ${\cal M}_2$. It is remarkable that 
the parametric uncertainty due to the input value of 
$\hat\alpha_s(M_Z)$ increases at the last step from 
$O(\hat\alpha_s^2)$ to $O(\hat\alpha_s^3)$ (see the last, 
purple error bars), while the central value for $\hat{m}_c 
(\hat{m}_c)$ remains stable. A closer look at the underlying 
formulae reveals that this is due to the bad convergence of the 
right-hand side of Eq.~(\ref{SR0}). It is therefore unlikely that 
knowledge of the next terms in the perturbative series will lead 
to a reduction of the error on the charm mass. Our final result 
for the charm quark mass,
\begin{equation}
\hat m_c(\hat m_c) = (1272 + 2616 \Delta \hat\alpha_s \pm 8)~{\rm MeV},
\label{cmresult}
\end{equation} 
is based on $({\cal M}_0, {\cal M}_2)$
where we explicitly exhibit its dependence on $\hat \alpha_s$ 
relative to our adopted central value\footnote{This central value 
   corresponds to $\hat\alpha_s(\hat m_c(\hat m_c)) = 0.383$,
   where we have used 4-loop running of $\hat\alpha_s$ including 
   the $b$ quark threshold effect using $\hat m_b(\hat m_b) = 4.2$ 
   GeV. We have cross-checked our implementation of the running of 
   the strong coupling constant against 
   {\sc Rundec}~\cite{Chetyrkin:2000yt} and found agreement at the 
   permille level.}, 
i.e., $\Delta \hat \alpha_s = \hat \alpha_s(M_z) - 0.1182$.


\begin{figure}[t!]
\begin{center}
\includegraphics[width=\textwidth]{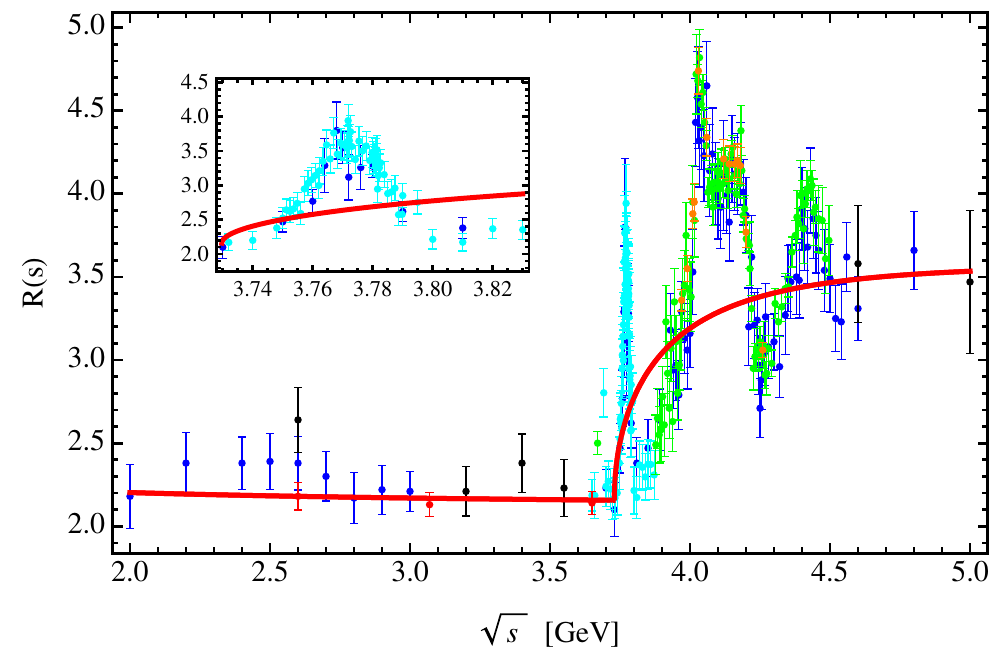}
\caption{
Data for the ratio $R(s)$ for $e^+e^- \rightarrow $ hadrons in the 
charm threshold region: Crystal Ball CB86 (green) 
\cite{Osterheld:1986hw}; BES00, 02, 06, 09 (black, blue, cyan, and red) 
\cite{Bai:1999pk,Bai:2001ct,Ablikim:2006mb,Ablikim:2009ad}, and CLEO09 
(orange) \cite{CroninHennessy:2008yi}. The full (red) curve is our 
parametrization of $R_c^{\rm cont}(s)$ with $\lambda_3^c = 1.23$ and 
$\hat{m}_c(\hat{m}_c) = 1.272$ GeV. The inner plot is a zoom into 
the energy range $2 M_{D^0} \leq \sqrt{s} \leq 3.83$~GeV.
}
\label{Rc}
\end{center}
\end{figure}

\section{Continuum data}
\label{Sec:data}

Our final result for the charm quark mass given above in 
Eq.~(\ref{cmresult}) is obtained from an analysis of sum rule moments 
without using experimental data from the continuum region. Instead, 
the continuum is described by the parameter $\lambda_3^c$, and 
determined simultaneously with $\hat{m}_c$ within our formalism.
In this section we show that Eq.~(\ref{ansatz}) indeed reproduces 
the experimental moments well (even though our ansatz must obviously 
fail to describe the underlying cross section data locally).

\begin{figure}[t!]
\begin{center}
\includegraphics[width=0.48\textwidth]{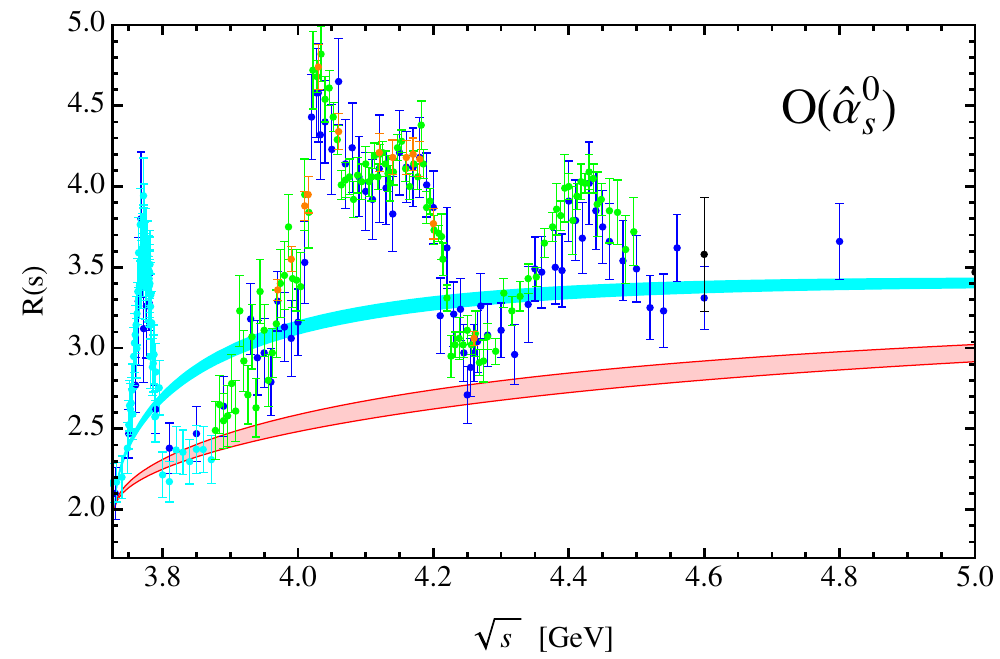}
\includegraphics[width=0.48\textwidth]{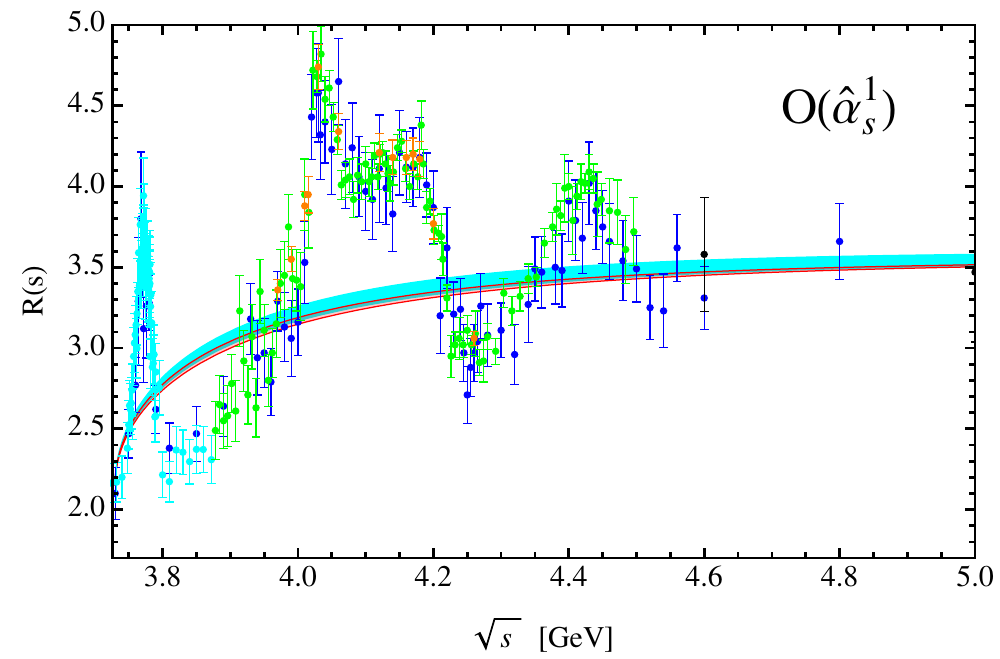} \\[1em]
\includegraphics[width=0.48\textwidth]{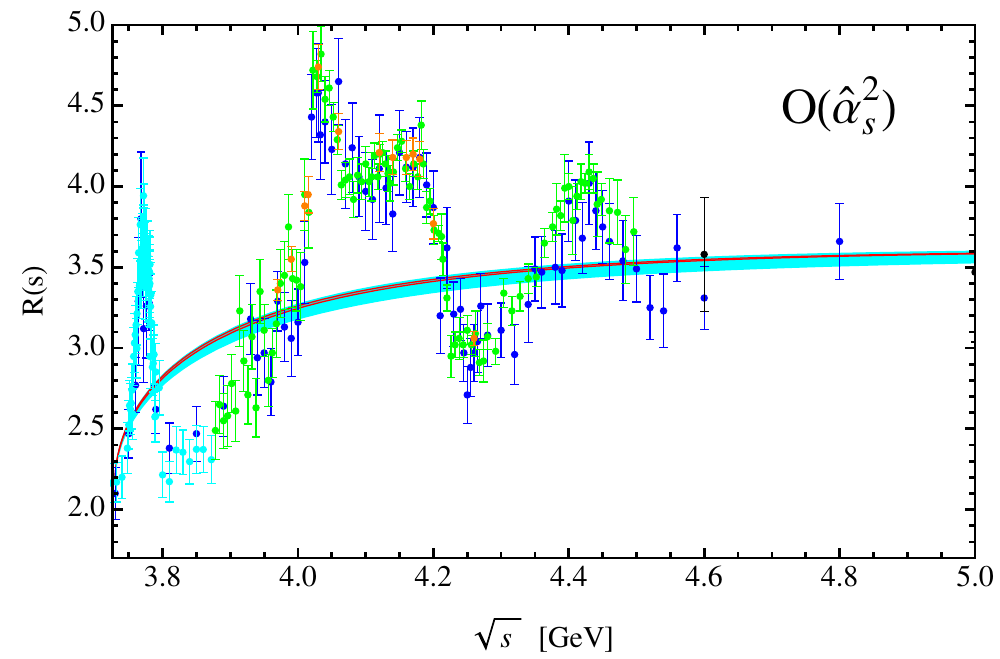}
\includegraphics[width=0.48\textwidth]{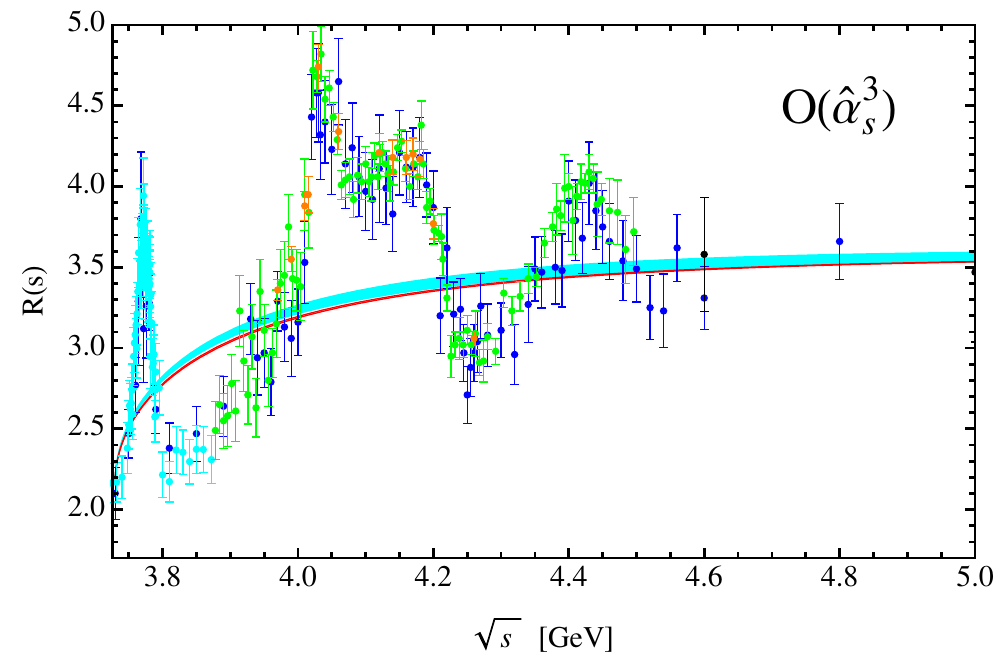}
\caption{
Same as Figure~\ref{Rc} at different orders of $\hat \alpha_s$.
For the full (red) curve we have used $\lambda_3^c$ determined 
from the pair of moments ${\cal M}_0$ and ${\cal M}_2$ at the 
respective order indicated in the plots with errors as described 
in the text, while for the blue band we have used 
$\lambda_3^{c,{\rm exp}}$ determined from data and their 
uncertainties at each order in $\hat \alpha_s$.  
}
\label{Rcan}
\end{center}
\end{figure}

In order to do so, we have to add the background from light-quark 
contributions \cite{Chetyrkin:1994js,Harlander:2002ur,Kuhn:2001dm, 
Kuhn:2007vp}, $R_{\rm background}(s) = R_{\rm uds}(s) + 
R_{\rm uds(cb)}(s) + R_{\rm singlet}(s) + R_{\rm QED}(s)$ to our 
model for the charm continuum. The first part, $R_{\rm uds}(s)$, 
is given by Eq.~(\ref{lambda1}) with $n_q=3$. Contributions from 
virtual or secondary heavy quarks, $R_{\rm uds(cb)}(s)$, appear 
first at ${\cal O}(\hat\alpha_s^2)$ and are known up to 
${\cal O}(\hat\alpha_s^3)$ in approximate form \cite{Harlander:2002ur}. 
In addition, at order ${\cal O}(\hat\alpha_s^3)$ one has to take 
into account the singlet contribution 
$R_{\rm singlet}(s)$~\cite{Harlander:2002ur} which 
is found to be numerically small, $-0.55 (\hat\alpha_s/\pi)^3$ 
\cite{Kuhn:2007vp}, but included for completeness. Finally, also 
corrections to QED have been taken into account through 
$R_{\rm QED}(s)$ \cite{Chetyrkin:1994js}. 

In Figure~\ref{Rc} we show the resulting $R$ ratio, $R(s) = 
R_c^{\rm cont}(s) + R_{\rm background}(s)$, compared with 
experimental $e^+ e^-$ annihilation data for final state hadrons. 
The data are compiled from Crystal Ball \cite{Osterheld:1986hw}, 
BES~\cite{Bai:1999pk,Bai:2001ct, Ablikim:2006mb,Ablikim:2009ad}, 
and CLEO \cite{CroninHennessy:2008yi}. The inset zooms into the 
region above threshold in the energy range of the $\Psi(3770)$ 
resonance, $2 M_{D^0} \leq \sqrt{s} \leq 3.83$~GeV, with data 
points from BES~\cite{Bai:2001ct,Ablikim:2006mb}. The full red 
curve in this figure is obtained from our parametrization of 
$R_c^{\rm cont}(s)$ in Eq.~(\ref{ansatz}), with $\lambda_3^c = 
1.23$ and $\hat{m}_c(\hat{m}_c)$ in Eq.\ (\ref{cmresult}).

\begin{table}[t]
\begin{center}
\footnotesize
{
\begin{tabular}{|c|c|c|c|c|c|c|}
\hline
\rule[-3mm]{0mm}{8mm}
Collab. & $n$ 
      & [$2 M_{D^0}, 3.872$] 
      & $[3.872,3.97]$ 
      & $[3.97,4.26]$
      & $[4.26,4.496]$
      & $[4.496,4.8]$
\\
\hline
\rule[-1mm]{0mm}{4mm}
\multirow{3}{*}{CB86} 
  & $0$ & --- & 0.0339(22)(24) & 0.2456(25)(172)  & 0.1543(27)(108) & --- 
  \\
  & $1$  &--- & 0.0220(14)(15) & 0.1459(16)(102) & 0.0801(14)(56) & --- 
  \\
  & $2$ &--- & 0.0143(9)(10) & 0.0868 (9)(61)& 0.0416(7)(29) & --- 
  \\[1mm]
\hline           
\rule[-1mm]{0mm}{4mm}
\multirow{3}{*}{BES02} 
   & $0$ & 0.0334(24)(17) & 0.0362(29)(18) & 0.2362(41)(118) & 0.1399(38)(70) & 0.1705(63)(85) 
   \\
   & $1$  &0.0232(17)(12) & 0.0235(19)(12) & 0.1401(24)(70) & 0.0726(20)(36)& 0.0788(30)(39) 
   \\
   & $2$  &0.0161(12)(8) & 0.0152(13)(8) & 0.0832(15)(42) & 0.0378(10)(19)& 0.0365(14)(18) 
   \\[1mm]
\hline
\rule[-1mm]{0mm}{4mm}
\multirow{3}{*}{BES06} 
   & $0$  &0.0311(16)(15) & --- & --- &--- & --- 
   \\
   & $1$  &0.0217(11)(11) &  --- & --- &--- & --- 
   \\
   & $2$  &0.0151(8)(7)     &  --- & --- &--- & --- 
   \\[1mm]
\hline
\rule[-1mm]{0mm}{4mm}
\multirow{3}{*}{CLEO09} 
   & $0$  &--- & --- & 0.2591(22)(52) &--- & --- 
   \\
   & $1$  &--- & --- &0.1539(13)(31) &--- & --- 
   \\
   & $2$  &--- & --- & 0.0915(8)(18) & --- & --- 
   \\[1mm]
\hline
\rule[-1mm]{0mm}{4mm}
\multirow{3}{*}{Total} 
   & $0$  &0.0319(14)(11) & 0.0350(18)(15) & 0.2545(18)(46) & 0.1448(27)(59) & 0.1705(63)(85) 
   \\
   & $1$  &0.0222(9)(8) & 0.0227(12)(10) & 0.1511(11)(27) & 0.0752(14)(31) & 0.0788(30)(39) 
   \\
   & $2$  &0.0155(6)(6) & 0.0147(8)(6) & 0.0899(6)(16) & 0.0391(7)(16) & 0.0365(14)(18) 
   \\[1mm]
\hline
\end{tabular}
}
\end{center}
\caption{
Contributions to the charm moments ($\times 10^{n}\, \mbox{GeV}^{2n}$) 
from different energy intervals (given in GeV) and experimental 
collaborations (CB86 \cite{Osterheld:1986hw}, BES02 \cite{Bai:2001ct}, 
BES06 \cite{Ablikim:2006mb}, CLEO09 \cite{CroninHennessy:2008yi}). The 
first errors are statistical and the second systematic. The last lines 
show the averaged results. 
}
\label{table:Intervals}
\end{table}

As described in the previous section, we can use data 
directly to obtain an experimental value $\lambda_3^{c, {\rm exp}}$. 
A comparison of our ansatz Eq.~(\ref{ansatz}) using either this 
experimental parameter or the one determined via pairs of moments 
as explained above is shown in Figure~\ref{Rcan}. Here we compare 
the resulting parametrizations of the charm continuum with data 
based on a determination of either $\lambda_3^c$ or $\lambda_3^{c, 
{\rm exp}}$ performed at different orders of pQCD. The red bands 
in Figure~\ref{Rcan} use $\lambda_3^c$ and $\hat{m}_c(\hat{m}_c)$ 
obtained at each order in $\hat \alpha_s$ while the blue bands use 
the same value for $\hat{m}_c(\hat{m}_c)$, but $\lambda_3^{c,{\rm exp}}$ 
instead, obtained using Eq.~(\ref{expmoment}) at each order in 
$\hat \alpha_s$. We observe that the differences when going from 
$O(\hat\alpha_s)$ to $O(\hat\alpha_s^2)$ and from $O(\hat\alpha_s^2)$ 
to $O(\hat\alpha_s^3)$ are of very similar size. This can be 
traced back to the slow convergence of the coefficients in 
$\lambda_1^c(s)$ of Eq.~(\ref{lambda1}). It is interesting to 
note that this does not lead to large changes in the values of 
$\hat m_c(\hat m_c)$ as is evident from Figure~\ref{Fig:cmass-an}. 
Note that we do not assign an extra error to $R_{\rm background}(s)$ 
since any uncertainty in this quantity is relegated to 
$\Delta \lambda_3^{c,{\rm exp}}$. 

We now have a closer look at the determination of moments 
from data in the energy range $2 M_{D^0} \leq \sqrt{s} \leq 4.8$~GeV. 
After light-quark background subtraction, we calculate the 
contributions to the moments in the considered energy range by 
numerical integration over the data. In Table~\ref{table:Intervals} 
we show the results for the zeroth, first, and second 
moments with data taken from Refs.~\cite{Osterheld:1986hw, 
Bai:2001ct,Ablikim:2006mb,CroninHennessy:2008yi}. We display 
the separate contributions from each individual collaboration split 
into five different energy intervals. These intervals have been chosen 
in such a way that none of the published data had to be split up 
into smaller segments than originally reported by the collaborations. 
The first errors are statistical and the second systematic. The statistical 
errors are uncorrelated. The systematic errors are taken uncorrelated 
among different collaborations, but completely correlated for data from 
the same collaboration. The last part in Table~\ref{table:Intervals}
(labelled 'Total') shows the average of the contributions from the 
different collaborations in each energy interval. The averaging procedure  
follows Ref.~\cite{Erler:2015nsa}. This prescription takes 
the relative weights of statistical and systematic errors from each 
collaboration into account. The breakdown of the total error 
$\Delta_{\rm tot}$ into statistical ($\Delta_{\rm stat}$) and 
systematic ($\Delta_{\rm sys}$) contributions can then be calculated 
in terms of the statistical ($\Delta {\rm C}^i_{\rm stat}$) and 
systematic ($\Delta {\rm C}^i_{\rm sys}$) errors of collaboration~$i$ 
contributing to the given interval as~\cite{Erler:2015nsa}
\begin{eqnarray}
(\Delta_{\rm stat})^2 &=&
\sum_i (\Delta {\rm C}^i_{\rm stat})^2 
\left( \frac{\Delta_{\rm tot}}
       {\sqrt{(\Delta {\rm C}^i_{\rm stat})^2 
       + (\Delta {\rm C}^i_{\rm sys})^2}} \right)^4,
\\
(\Delta_{\rm sys})^2 &=&
\sum_i (\Delta {\rm C}^i_{\rm sys})^2 
\left( \frac{\Delta_{\rm tot}}
       {\sqrt{(\Delta {\rm C}^i_{\rm stat})^2 
       + (\Delta {\rm C}^i_{\rm sys})^2}} \right)^4.
\end{eqnarray}

\begin{table}[t]
\begin{center}
\begin{tabular}{|c|ccc|}
\hline
\rule[-3mm]{0mm}{9mm}
$n$ & Data & $\lambda^c_3 = 1.34(17)$ & $\lambda^c_3 = 1.23$ \\
\hline
\rule[-2mm]{0mm}{7mm}
0 & 0.6367(195) & 0.6367(195) & 0.6239 \\
\rule[-2mm]{0mm}{6mm}
1 & 0.3500(102)  & 0.3509(111)  & 0.3436 \\
\rule[-2mm]{0mm}{6mm}
2 & 0.1957(54)  & 0.1970(65)  & 0.1928 \\
\rule[-2mm]{0mm}{6mm}
3 & 0.1111(29)  & 0.1127(38)  & 0.1102 \\
\rule[-2mm]{0mm}{6mm}
4 & 0.0641(16)  & 0.0657(23)  & 0.0642 \\
\rule[-2mm]{0mm}{6mm}
5 & 0.0375(9)   & 0.0389(14)   & 0.0380 \\
\hline
\end{tabular}
\end{center}
\caption{
Contributions to the charm moments ($\times 10^{n}\, \mbox{GeV}^{2n}$) 
from the energy range $2 M_{D^0} \leq \sqrt{s} \leq 4.8$~GeV. For 
the results in the column labeled 'Data', light-quark contributions 
have been subtracted using the pQCD prediction at order ${\cal O} 
(\hat\alpha_s^3)$. These entries are obtained from the data displayed 
in Table~\ref{table:Intervals}, taking into account the correlation 
of systematic errors within each experiment. The third column uses 
$\hat{m}_c = 1.272\, {\rm GeV}$ and $\lambda_3^{c,{\rm exp}}$ 
determined by the zeroth experimental moment (see text for details). 
The last column shows the theoretical prediction for the moments 
using $\hat{m}_c = 1.272\, {\rm GeV}$ and $\lambda^c_3 = 1.23$, not 
including condensates. 
}
\label{table:Data1}
\end{table}

The final averaged result for the energy range between $2 M_{D^0}$ 
and $\sqrt{s} = 4.8$~GeV is given in Table~\ref{table:Data1} up to 
the fifth moment. The errors shown there are the combined statistical 
and systematic ones. The third column shows the theoretical moments, 
again restricted to the region $2 M_{D^0} \leq \sqrt{s} \leq 4.8$ GeV, 
using $\hat{m}_c = 1.272$ GeV as input, and choosing the parameter 
$\lambda_3^c$ to match the zeroth experimental moment, 
\be 
  \int_{(2M_{D^0})^2}^{(4.8{\rm \, GeV})^2} 
  \frac{{\rm d} s}{s} R_c^{\rm cont}(s) 
  \Big|_{\hat{m}_c = 1.272 \, {\rm GeV}} 
  = 
  {\cal M}_0^{\rm Data} = 0.6367(195) 
\label{expmoment}
\ee
which gives the value $\lambda_3^{c,{\rm exp}} = 1.34(17)$. 
Finally, the fourth column shows the theoretical calculation 
of the moments in the threshold region for $\hat{m}_c = 1.272$~GeV 
and using the previous value $\lambda_3^c = 1.23$ found above. 
The agreement between these three columns is remarkable and shows 
that the data in this restricted energy range is well described 
by $R_c^{\rm cont}(s)$ in Eq.\ (\ref{ansatz}). 

In the analysis of the previous section only the two narrow resonances 
below the heavy-quark threshold, i.e., the $J/\Psi$ and $\Psi(2S)$, 
have been treated separately, while the other charmonium states were 
included in the continuum. We now study the effect of treating also 
the $\Psi(3770)$ resonance separately and include it based on the 
narrow width approximation. 

\begin{table}[t]
\begin{center}
\begin{tabular}{|c|ccc|}
\hline
\rule[-3mm]{0mm}{9mm}
$n$ & Data & $\Psi(3770)\big|_{\rm BW}$ & $\lambda^c_3 = 1.23$\\
\hline
\rule[-2mm]{0mm}{7mm}
0 
& 0.0268(14) 
& 0.0291(20) 
& 0.0339 \\
\rule[-2mm]{0mm}{6mm}
1 
& 0.0188(10) 
& 0.0204(14) 
& 0.0236 \\
\rule[-2mm]{0mm}{6mm}
2
& 0.0132(7)
& 0.0143(10)
& 0.0165 \\
\rule[-2mm]{0mm}{6mm}
3
& 0.0092(5)
& 0.0101(7)
&  0.0115 \\
\rule[-2mm]{0mm}{6mm}
4
& 0.0065(3)
& 0.0071(5)
& 0.0080 \\
\rule[-2mm]{0mm}{6mm}
5 
& 0.0045(2)
& 0.0050(3)
& 0.0056 \\
\hline
\end{tabular}
\end{center}
\caption{
Contributions to the charm moments ($\times 10^{n}\, \mbox{GeV}^{2n}$) 
from the energy range $2 M_{D^0} \leq \sqrt{s} \leq 3.83$~GeV. For 
the results in the column labeled 'Data', light-quark contributions 
have been subtracted using the pQCD prediction at order  ${\cal O} 
(\hat\alpha_s^3)$. 'BW' refers to the Breit-Wigner expression in 
Eq.~(\ref{BW}). The last column shows the theoretical prediction for 
the moment using $\hat{m}_c = 1.272\, {\rm GeV}$ and $\lambda^c_3 = 
1.23$, not including condensates. 
}
\label{table:Data2}
\end{table}

Table~\ref{table:Data2} shows a comparison of charm moments 
in the closer vicinity of the $\Psi(3770)$ resonance. In the 
column labeled 'Data', the moments have been calculated as above, 
i.e.\ directly from data, but now restricted to the energy range 
$2 M_{D^0} \leq \sqrt{s} \leq 3.83$ GeV (see the inner plot of 
Figure~\ref{Rc}) again after subtraction of the light-quark 
contribution. The upper limit of this energy interval is chosen 
to cover the $\Psi(3770)$ resonance completely. In the third 
column (labeled $\Psi(3770)$) we display results obtained 
assuming that the resonance is not narrow enough to justify 
using Eq.~(\ref{Rres}). Instead, we use the full Breit-Wigner form,
\be\label{BW}
  R^{\rm \Psi(3770)}_{\rm BW} 
  = 
  \frac{9 \Gamma^e_R \Gamma_R}{\alpha_{\rm em}^2(M_R^2)} \,  
  \frac{M_R^2}{(s-M_R^2)^2+ \Gamma_R^2 M_R^2}
\ee
with  $M_{\Psi(3770)} = 3.773$ GeV, $\Gamma_{\Psi(3770)}^e = 
0.262(18)$~keV, and $\Gamma_R = 27.2 \pm 1.0$~MeV for the total 
width of the $\Psi(3770)$~\cite{Agashe:2014kda}. Even though 
one can not assume that the $\Psi(3770)$ resonance saturates 
the charm cross section in this narrow energy range, a 
Breit-Wigner description agrees with the prediction based on 
Eq.~(\ref{ansatz}) when $R_c^{\rm cont}(s)$ is used with 
$\lambda_3^c = 1.23$. 

Another way to estimate the impact of the $\Psi(3770)$ 
region on the charm quark mass is to include the 
$\Psi(3770)$ into our parameterization for $R_c^{\rm Res}(s)$ as a 
$\delta$-function~(\ref{Rres}) and re-evaluate $\hat{m}_c(\hat{m}_c)$ 
and $\lambda^c_3$ using the pair of moments $({\cal M}_0,{\cal M}_2)$. 
We find shifts of $\Delta \hat{m}_c(\hat{m}_c) = -2.4$ MeV and 
$\Delta \lambda^c_3 = -0.12$. This procedure adds more weight at 
lower $\sqrt{s}$; therefore the 2nd moment will increase relative 
to the 0th moment. As a compensation, the charm mass will decrease. 
The uncertainties associated with the $\Psi(3770)$ resonance 
parameters are negligible.

\begin{table}[t]
\begin{center}
\begin{tabular}{|c|ccc|}
\hline
\rule[-3mm]{0mm}{9mm}
$n$ & Data & $\lambda^c_3 = 1.15(16)$ & $\lambda^c_3 = 1.23$ \\
\hline
\rule[-2mm]{0mm}{7mm}
0 & 0.6149(189) & 0.6149(189) & 0.6239 \\
\rule[-2mm]{0mm}{6mm}
1 & 0.3378(98)  & 0.3385(108)  & 0.3436 \\
\rule[-2mm]{0mm}{6mm}
2 & 0.1886(52)  & 0.1898(63)  & 0.1928 \\
\rule[-2mm]{0mm}{6mm}
3 & 0.1070(28)  & 0.1085(37)  & 0.1102 \\
\rule[-2mm]{0mm}{6mm}
4 & 0.0616(16)  & 0.0631(22)  & 0.0642 \\
\rule[-2mm]{0mm}{6mm}
5 & 0.0360(9)  & 0.0374(14) & 0.0380 \\
\hline
\end{tabular}
\end{center}
\caption{
Contributions to the charm moments ($\times 10^{n}\, \mbox{GeV}^{2n}$) 
from the energy range $2 M_{D^0} \leq \sqrt{s} \leq 4.8$~GeV evaluated 
after taking into account a $2\,\%$ shift of the normalization of the 
pQCD prediction to fit the data, to be compared with Table 
\ref{table:Data1}. In this case we obtained $\hat{m}_c(\hat{m}_c) 
= 1.272$~GeV and $\lambda^c_3 
= 1.15(16)$ after matching with the zeroth experimental moment.
}
\label{table:Data1a}
\end{table}

One may also raise the question whether the specific prescription 
for the subtraction of the light-quark background in the charm 
sub-threshold region should be modified, or whether an additional 
error should be assigned to this prescription. To check the impact 
of this effect, we have fitted the pQCD prediction including a free 
normalization factor to the data between $\sqrt{s} = 2$ GeV and 
$2M_D$. We found a normalization factor of $1.02 \pm 0.01$ and the 
moments changed to the values given in Table~\ref{table:Data1a}. 
The charm mass is shifted by less than 1~MeV if this modified 
light-quark subtraction is used instead of the prescription 
described above. 


\section{Alternative fits}
\label{Sec:errors}

The approach described in Section \ref{Sec:formalism} 
requires to make a choice of a pair of moments from which to 
determine a value for the charm mass. The results displayed 
in Table~\ref{Tab:MomentsBudget} (see the line labelled 'Total') 
show that the pair $({\cal M}_0,{\cal M}_2)$ leads to the smallest 
error (not taking into account the uncertainty from $\hat \alpha_s$). 
A more careful examination reveals that different sources of the 
errors contribute with different weights. While the pair 
$({\cal M}_0,{\cal M}_1)$ is the least sensitive to the gluon 
condensate and most strongly affected by the uncertainty of the 
experimental data from the continuum region, the pair $({\cal M}_0, 
{\cal M}_3)$ has little sensitivity to the continuum data and is 
more strongly affected by the truncation of the OPE. In this 
section we investigate the possibility to perform a fit using also 
more than two moments. In addition, the approach that we discuss 
now will allow us to take into account uncertainties on the 
parameters entering the moments in a more systematic way. This will 
confirm the robustness of our main result in Eq.~(\ref{cmresult}).

\begin{table}[t!]
\begin{center}
\begin{tabular}{|c|r|r|r|r|}
\hline
\rule[-3mm]{0mm}{9mm}
& Constraints 
& $({\cal M}_0, {\cal M}_1, {\cal M}_2)_\rho$ 
& ${\cal M}_0,~({\cal M}_1, {\cal M}_2)_\rho$ 
& ${\cal M}_0,~({\cal M}_1, {\cal M}_2, {\cal M}_3)_\rho$ 
\\[5pt]
\hline
\rule[-3mm]{0mm}{9mm}
$\rho$ 
& 
& $-0.06$ 
& $-0.05$
& 0.32
\\[5pt]
\hline 
\rule[-3mm]{0mm}{9mm}
$\hat m_c(\hat m_c)$~[GeV] 
& 
& $1.275(8)$ 
& $1.275(8)$ 
& $1.271(7)$ 
\\[5pt]
$\lambda_3^c$ 
& 
& $1.19(8)$ 
& $1.19(8)$  
& $1.19(7)$  
\\[5pt]
$\Gamma^e_{J/\Psi}$~[keV] 
& $5.55(14)$ 
& $5.57(14)$
& $5.57(14)$
& $5.59(14)$ 
\\[5pt]
$\Gamma^e_{\Psi(2S)}$~[keV] 
& $2.36(4)$ 
& $2.36(4)$ 
& $2.36(4)$ 
& $2.36(4)$ 
\\[5pt]
$C_G$~[GeV$^4$] 
& $0.005(5)$ 
& $0.005(5)$ 
& $0.005(5)$ 
& $0.004(5)$ 
\\[5pt]
$\hat\alpha_s(M_z)$ 
& $0.1182(16)$ 
& $0.1178(15)$
& $0.1178(15)$ 
& $0.1173(15)$ 
\\[5pt]
\hline
\end{tabular}
\end{center}
\caption{
Fit results using the first three (four) moments with different 
scenarios for the correlation between the moments (see text). 
}
\label{Tab:Tfits}
\end{table}

We define a $\chi^2$-function by 
\begin{equation}
\chi^2 = 
\frac{1}{2} \sum_{n,m} 
\left( {\cal M}_n - {\cal M}_n^{\rm pQCD} \right) 
\left({\cal C}^{-1}\right)^{nm} 
\left( {\cal M}_m - {\cal M}_m^{\rm pQCD} \right) 
+ \chi^2_c
\label{chi2}
\end{equation}
where the ${\cal M}_n$ are the sum rule expressions for the moments 
defined in Eq.\ (\ref{SRder}) for $n>0$, and Eq.\ (\ref{SR0}) 
multiplyed by $3Q_c^2$ for $n=0$. The corresponding predictions 
from perturbative QCD, ${\cal M}_n^{\rm pQCD}$, were defined in 
Eq.\ (\ref{Mth}). We will consider different fit scenarios where 
we include different sets of moments in the sum of Eq.\ 
(\ref{chi2}). The coefficients $\left({\cal C}^{-1}\right)^{nm}$ 
are the elements of a correlation matrix which we choose as 
\begin{equation}
{\cal C} 
= 
\frac{1}{2} \sum_{n,m} 
\rho^{|n-m|} \Delta {\cal M}_n^{(3)} \Delta {\cal M}_m^{(3)} 
\end{equation}
with the truncation error $\Delta {\cal M}_m^{(3)}$ defined in 
Eq.\ (\ref{truncerror}). The factor $\rho^{|n-m|}$ serves as an 
{\em ansatz\/} to include correlations between different moments, 
where $\rho$ itself is a free parameter. In addition, we will also 
consider cases with vanishing correlations. In particular, the 
correlation between the zeroth and the other moments may be much 
smaller than the correlations between the higher moments (especially 
adjacent ones). The term $\chi^2_c$ in Eq.\ (\ref{chi2}) imposes 
constraints on the variation of parameters like the electronic 
widths in Table~\ref{ResPDG}, the strong coupling constant, 
$\hat\alpha_s(M_z)^{\rm exp} = 0.1182 \pm 0.0016$~\cite{Agashe:2014kda},
and the gluon condensate, $C_G^{\rm exp} = 0.005 \pm 
0.005$~GeV$^4$~\cite{Dominguez:2014fua},
$$
\chi^2_c =
\left[\frac{\Gamma^{e}_{J/\Psi} - \Gamma^{e,{\rm exp}}_{J/\Psi}}%
{\Delta \Gamma^{e}_{J/\Psi}}\right]^2 + 
\left[\frac{\Gamma^{e}_{\Psi(2S)} - \Gamma^{e,{\rm exp}}_{\Psi(2S)}}%
{\Delta \Gamma^{e}_{\Psi(2S)}}\right]^2 + 
\left[\frac{\hat\alpha_s(M_z) - \hat\alpha_s(M_z)^{\rm exp}}%
{\Delta\hat\alpha_s(M_z)}\right]^2 + 
\left[\frac{C_G - C_G^{\rm exp} }%
{\Delta_G}\right]^2.
$$
Thus, $\chi^2$ is a function of six parameters, 
$\hat m_c(\hat m_c)$, $\lambda_3^c$ and $\hat\alpha_s$ for the 
continuum part of the moments, the electronic widths 
$\Gamma^e_{J/\Psi}$ and $\Gamma^e_{\Psi(2S)}$ for the 
resonance contributions, and $C_G$. In addition there is the 
correlation parameter $\rho$. 

\begin{table}[t!]
\rule[0mm]{0mm}{3mm}
\begin{center}
\begin{tabular}{|c|c|c|c|}
\hline
\rule[-3mm]{0mm}{9mm}
& ${\cal M}_0,~{\cal M}_1$ 
& ${\cal M}_0,~{\cal M}_2$ 
& ${\cal M}_0,~{\cal M}_3$ 
\\[5pt]
\hline
\rule[-3mm]{0mm}{9mm}
$\hat m_c(\hat m_c)$~[GeV] 
& $1.281(9)$ 
& $1.272(7)$ 
& $1.269(5)$ 
\\[5pt]
$\lambda_3^c$ 
&$1.15(5)$ 
& $1.23(6)$ 
& $1.26(8)$ 
\\[5pt]
\hline
\end{tabular}
\end{center}
\caption{
Fit results using pairs of moments without correlation. 
}
\label{Tab:Tfits2}
\end{table}


We now determine all or a subset of the parameters by minimzing 
$\chi^2$. First, we observe that without correlations between 
the moments, the minimal $\chi^2$ is small. This is likely
an indication that the theoretical moments are indeed 
correlated. Taking into account correlations as described 
above, we now determine the correlation parameter $\rho$ from 
the condition that $\chi^2_{\rm min}$ be equal to the 
number of degrees of freedom for each considered fit scenario. 
Then we determine the allowed parameter ranges by solving the 
equation $\chi^2 = \chi^2_{\rm min} + 1$. 

In Table~\ref{Tab:Tfits} we collect the results of three typical 
fit scenarios based on the first three (four) moments.
Columns three, four and five are results of 6-parameter fits. 
In the first case (column 3) we assume that all three moments 
are correlated as described above. In the second case (column 4) 
we assume that only the first and the second moments are correlated. 
For the results in column 5 we have also included the moment 
${\cal M}_3$. In this case the uncertainty is slightly smaller, but
we have more confidence in lower moments which are less sensitive to 
the condensates. Moreover, in this scenario the correlation is larger. 
The errors are obtained by projecting the contour $\chi^2 = 
\chi^2_{\rm min} + 1$ onto each of the six parameters considered. 
We observe that the results for the charm mass as well as for the 
continuum parameter $\lambda_3^c$ are stable. The other four 
parameters are shifted away from their experimental values only 
slightly and stay well within their uncertainties. We have performed 
additional fits in modified scenarios, e.g., using the first three 
and four moments without correlations. We observe only small shifts 
in the fitted charm mass of not more than 4~MeV, which is well 
within our quoted uncertainty.  Also fits with only two moments are 
stable and the results for $\hat m_c$ and $\lambda_3^c$ agree 
within errors. A few typical examples are shown in 
Table~\ref{Tab:Tfits2} where no correlation is assumed. 

\begin{table}[t!]
\begin{center}
\begin{tabular}{|l|c|c|}
\hline
\rule[-3mm]{0mm}{8mm}
 & 
 $\Delta \hat{m}_c(\hat{m}_c) [\mbox{MeV}]$ &
 $\Delta \hat{m}_c(\hat{m}_c) [\mbox{MeV}]$
\\
\hline 
\rule[-2mm]{0mm}{7mm}
Central value & 
1274.5  &
1272.4
\\
\rule[-2mm]{0mm}{6mm}
$\Delta \Gamma^e_{J/\Psi}$ & 
5.9 &
4.5
\\
\rule[-2mm]{0mm}{6mm}
$\Delta \Gamma^e_{\Psi(2S)}$ & 
1.4 &
0.4
\\
\rule[-2mm]{0mm}{6mm}
Truncation & 
$-$ &
5.9
\\
\rule[-2mm]{0mm}{6mm}
$\Delta \lambda_3^c$ & 
3.0 &
2.3
\\
\rule[-2mm]{0mm}{6mm}
Condensates 
& 
1.1 &
1.9
\\
\rule[-2mm]{0mm}{6mm}
$\Delta \hat\alpha_s(M_Z)$ & 
5.4 &
4.2
\\
\hline
\rule[-2mm]{0mm}{7mm}
Total &
8.7 &
9.0
\\
\hline
\end{tabular}
\end{center}
\caption{
Error budget for the fit scenario ${\cal M}_0,~({\cal M}_1, 
{\cal M}_2)_\rho$ from one-parameter fits for the parameter 
displayed in each line and using the other parameters fixed 
at their values given in the last column of Table~\ref{Tab:Tfits}. 
The last column compares this with our main result determined from 
the pair $({\cal M}_0, {\cal M}_2)$ in Table~\ref{Tab:MomentsBudget}.
}
\label{Tab:errorbudget}
\end{table} 

Table~\ref{Tab:errorbudget} contains the error budget for the 
uncertainty of the charm mass in the fit scenario using the 
moments for $n=0$, 1, and 2 with a correlation between ${\cal M}_1$ 
and ${\cal M}_2$. The errors are from one-parameter fits with all 
other parameters fixed at their best fit values, given in the 
fourth column of Table~\ref{Tab:Tfits}. 


\section{Comparison with previous work}

Our result agrees within errors with the previous 
analysis~\cite{Erler:2002bu} where $\hat m_c(\hat m_c) 
= 1.289^{+0.040}_{-0.045}$~GeV was found. We traced the 
moderate numerical difference to the following individual shifts:
\begin{itemize}
\item
The resonance parameters have been updated in recent years. 
In particular, the value of $\Gamma^e_{J/\Psi}$ changed from 
$5.26(37)$~keV to $5.55(14)$~keV. The larger electronic widths 
increase the resonance contribution to the higher moments
more than to the zeroth moment. As a consequence of the new 
values of $\Gamma^e_R$ the charm quark mass is shifted by 
$-12$ MeV.
\item 
Both the central value and the uncertainty of $\hat\alpha_s(M_z)$ 
have changed. In Ref.~\cite{Erler:2002bu}, $\hat\alpha_s(M_z) = 
0.1221$ was used, while in the current work we use 
$\hat\alpha_s(M_z) = 0.1182$. For $\hat m_c(\hat m_c)$ this 
leads to a shift of $-10$ MeV.
\item
In the current work, we use a non-zero central value for the 
contribution from the gluon condensate. $C_G = 0.005$~GeV$^4$ 
with $\hat\alpha_s(M_z) = 0.1221$ induces a shift of the charm 
quark mass of $-2$ MeV. 
\item
Terms proportional to $\pi^2$ (i.e., $\zeta(2)$) in the 
definition of $\lambda_1^c(s)$ in \cite{Erler:2002bu} are in 
fact not present. Correcting this leads to a shift of $+3$~MeV 
for $\hat m_c(\hat m_c)$.
\item
Progress in theoretical calculations of higher-order terms 
allows us to include effects of ${\cal O}(\hat\alpha_s^3)$ 
in the perturbative coefficients of the moments, in the
definition of $\lambda_1^c(s)$, and in the running of 
$\hat\alpha_s$. The combined effect of all 
${\cal O}(\hat\alpha_s^3)$ terms induces a shift of $+4$~MeV.
\item 
There are far more data for the cross section of $e^+ e^-$ 
annihilation in the charm threshold region, as well as in the 
region below the charm threshold. These new data  
affect only the calibration of the uncertainty, however. 
\item 
The inclusion of the singlet piece in the light-quark background 
subtraction affects the charm mass only at the sub-MeV level.
\item 
Shifts induced by possible variations in $\hat m_b(\hat m_b)$ 
amount to at most a few hundred keV. 
\end{itemize}
We summarize the various shifts in the charm quark mass 
in Table~\ref{Tab:comparison} where we also include the 
corresponding shifts in the continuum parameter $\lambda_3^c$. 

\begin{table}[t!] 
\begin{center}
\begin{tabular}{|c|c|c|}
\hline
\rule[-3mm]{0mm}{9mm}
& 
$ \hat m_c(\hat m_c)$ & $\lambda_3^c$ 
\\
\hline
\rule[-2mm]{0mm}{7mm}
Ref.~\cite{Erler:2002bu} & $1289$ MeV & $1.97$
\\
\hline
\rule[-2mm]{0mm}{7mm}
Resonances & $-12$ MeV & $-0.16$
\\
\rule[-2mm]{0mm}{7mm}
$\hat\alpha_s(M_z)$ & $-10$ MeV & $-0.23$ 
\\
\rule[-2mm]{0mm}{7mm}
Gluon condensate & $-2$ MeV & $+0.02$ 
\\
\rule[-2mm]{0mm}{7mm}
No $\pi^2$ terms & $+3$ MeV & $-0.36$ 
\\
\rule[-2mm]{0mm}{7mm}
${\cal O}(\hat\alpha_s^3)$ & $+4$ MeV & $-0.57$
\\
\hline
\end{tabular}
\caption{
Comparison between Eq.\ (\ref{cmresult}) and the result of 
Ref.~\cite{Erler:2002bu} where $\hat\alpha_s(M_z) = 0.1221$ was 
used and the breakdown of the different sources of errors. 
See the text for details. 
}
\label{Tab:comparison}
\end{center} 
\end{table} 

We also find agreement within errors with other charm quark 
mass determinations in the literature. The result of 
Ref.~\cite{Chetyrkin:2010ic}, $\hat m_c(\hat m_c)=1.279(13)$~GeV, 
obtained from the first moment and using $\hat \alpha_s(M_z) = 
0.1189$ and $C_G = 0.006(12)$~GeV$^4$, agrees very well with 
the value we find from the pair $({\cal M}_0, {\cal M}_1)$ in 
Table~\ref{Tab:MomentsBudget}. Ref.~\cite{Bodenstein:2011ma} 
used the second moment and a modification of the kernel of the 
sum rule in Eq.~(\ref{SR}) to obtain $\hat m_c(\hat m_c) = 
1.278(9)$~GeV using as input $\hat \alpha_s(M_z) = 0.1184(7)$ 
and $C_G = 0.01(1)$~GeV$^4$. Ref.~\cite{Narison:2011xe} is based 
on higher moments extracting $C_G = (0.022 \pm 0.004)$~GeV$^4$ 
which then yields $\hat m_c(\hat m_c) = 1.261(16)$~GeV. While 
the charm production cross section is described by a set of 
six narrow-width resonances, combined with a continuum from 
perturbative QCD, the difference between Ref.~\cite{Narison:2011xe} 
and our result can be explained to a large extent by the 
different values of the gluon condensate. A more recent 
analysis~\cite{Dehnadi:2015fra} extracted the charm quark mass 
from the first moment sum rule using experimental data from a 
previous analysis \cite{Dehnadi:2011gc}. We can reproduce their 
final result, $\hat m_c(\hat m_c) = 1.288(20)$~GeV, if we use 
their input values, $\hat \alpha_s(M_z) = 0.1184(21)$ and 
$C_G = 0.006(12)$~GeV$^4$, and the electronic widths of the 
$\Psi(2S)$ resonance before its recent update. The authors of 
Ref.~\cite{Dehnadi:2011gc} have also discussed the $Z$-boson 
contribution to the QCD moments and found it to be very small. 
This contribution is neglected in our work.

\begin{figure}[t]
\begin{center}
\includegraphics[width=0.9\textwidth]{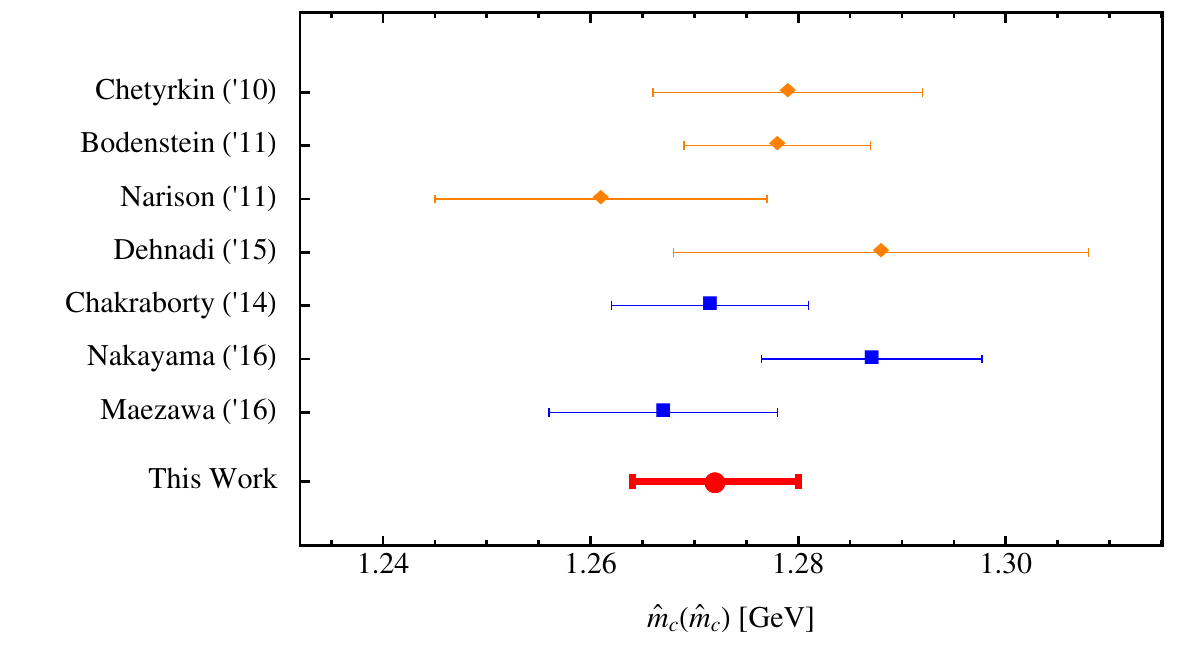}
\caption{
Recent charm quark mass determinations (see 
text for details and references). 
}
\label{Fig:Comparison}
\end{center} 
\end{figure}

Beyond the phenomenological charm quark mass determinations 
discussed so far, there are also precise lattice QCD calculations.
Ref.~\cite{Chakraborty:2014aca} reports $\hat m_c(\hat m_c) 
= 1.2715(95)$~GeV with $\hat \alpha_s(M_z) = 0.11822(74)$ from
a study of pseudoscalar-pseudoscalar correlators. 
Ref.~\cite{Nakayama:2016atf}, using M\"obius domain-wall 
fermions to describe both light and charm quarks, obtained 
the value $\hat m_c(\hat m_c) = 1.2871(106)$~GeV for 
$\hat \alpha_s(M_z) = 0.1177(25)$. A value for the gluon 
condensate was also fitted and found to be compatible with 
zero. Finally, Ref.~\cite{Maezawa:2016vgv} used the Highly 
Improved Staggered Quark action and moments of the 
pseudsocalar-charmonium correlator, and found $\hat m_c(\hat m_c) 
= 1.267(11)$~GeV and $\hat \alpha_s(M_z) = 0.1162(75)$. 
These results are displayed in Figure~\ref{Fig:Comparison}. 


\section{Conclusions}
\label{Sec:conclusions}

We presented a determination of the charm quark mass from 
QCD sum rules with a careful determination of the uncertainty. 
We included the zeroth moment in our strategy and exploited the 
consistency between different moments. The usefulness of the 
zeroth moment is not its sensitivity to $\hat m_c$ itself 
(which is mild), but to the continuum contribution to which 
it is very sensitive. While the coefficient $A_3$ entering the 
zeroth moment through Eq.\ (\ref{SR0}) is uncomfortably large,
we fully account for this by our conservative error estimate 
in Eq.~(\ref{truncerror}). It is interesting to note, however, 
that the dispersive calculation of the value of the electromagnetic 
coupling $\alpha_{\rm em}$ at the $Z$-pole (an essential input into 
the calculation of the $W$ boson mass, as well as for the electroweak 
program at LEP and any future $Z$ factory), effectively relies 
on precisely the zeroth moment. One usually neglects to make any 
reference to what value of $\hat m_c$ the charm quark contribution 
to $\alpha_{\rm em}(M_Z)$ would correspond to and whether that 
value is compatible with other determinations. It is this relation 
that shows poor convergence. Our approach amounts to determining 
the error on $\hat m_c$ by demanding compatibility with some higher 
moment. Our value of $\hat m_c$ in Eq.~(\ref{cmresult}) can then be 
used along the lines of Ref.~\cite{Erler:1998sy}, where perturbative 
QCD was used to directly compute $\alpha_{\rm em}(M_Z)$ in the 
$\overline{\rm MS}$ scheme (with $\hat m_c$ as an external input).

Note, that at this level of accuracy a subtle issue arises:
we have implicitly defined $\hat m_c(\hat m_c)$ as the 
$\overline{\rm MS}$ mass w.r.t.\ both QCD and QED, as can be 
seen e.g.\ from Eq.~(\ref{Cnth}). Alternatively adopting a 
hybrid definition, $\hat m_c(\hat m_c)^{\rm (pole-QED)}$, 
which assumes the on-shell mass w.r.t.\ QED so that the 
QED treatment of quarks mirrors more closely that of leptons, 
one has,
\be
\hat m_c(\hat m_c)^{\rm (pole-QED)} 
= 
\hat m_c(\hat m_c) 
\left[ 1+ Q_c^2 \frac{\alpha_{em}(m_c)}{\pi} 
+ {\cal O}(\alpha_{em}^2) \right] 
= (1274 + 2616 \Delta \hat\alpha_s \pm 8)~\mbox{MeV}.
\ee

The largest contribution to the total uncertainty in the charm 
quark mass is the truncation error. Since the coefficients of 
the perturbative expansion of the moments exhibit no 
compelling evidence for a convergent behaviour, we consider 
it unlikely that knowledge of the next terms in the perturbative 
series will lead to a reduction of the error on the charm mass. 

In future work we will apply our approach also to the case 
of the bottom quark. We believe that we should be able to 
justify an error estimate for $\hat m_b(\hat m_b)$ of about 15~MeV. 
This follows from the fact that our determination of the charm 
mass scales with the difference between half of the lowest-mass 
resonance, $M_{J/\Psi}/2$, and the quark mass $\hat m_c(\hat m_c)$.
Our uncertainty of $\Delta \hat m_c(\hat m_c) = 8$~MeV corresponds 
to $2.9\,\%$ of this difference. Translated to the case of the 
$b$ quark, using $M_{\Upsilon} = 9.460$~GeV and $\hat m_b(\hat m_b) 
\sim 4.200$~GeV, this would correspond to 15~MeV.


\section*{Acknowledgments}

We thank Irinel Caprini and Konstantin Chetyrkin for helpful 
discussions. 
This work has been partially supported by DFG through the 
Collaborative Research Center ``The Low-Energy Frontier of 
the Standard Model'' (SFB~1044). 
JE is supported by PAPIIT (DGAPA--UNAM) project IN106913 and 
by CONACyT (M\'exico) project 252167--F and acknowledges financial 
support from the Mainz cluster of excellence PRISMA. 


\end{document}